\documentclass[12pt]{article} 
\pdfoutput=1
\usepackage[T1]{fontenc}
\usepackage{amsmath}
\usepackage{amssymb}
\usepackage{amsthm}
\usepackage{epsfig}
\usepackage{graphicx}
\newtheorem{thm}{Theorem}
\newtheorem{cor}[thm]{Corollary}

\def\singlespace{\baselineskip=\normalbaselineskip}
\def\doublespace{\baselineskip=\normalbaselineskip \multiply\baselineskip by 7
\divide\baselineskip by 5}

\begin{document}

\doublespace

\thispagestyle{empty}
\begin{center}
{\Large\bf Computationally Efficient Nonparametric Importance Sampling}
 \vskip 5mm
Jan C.\ Neddermeyer\footnote{University of Heidelberg, Institute of Applied Mathematics, Im
Neuenheimer Feld 294, D-69120 Heidelberg, Germany.}

\end{center}

\vskip 5mm

\doublespace
\noindent
\emph{Extended Abstract.}
The variance reduction established by importance sampling strongly depends on
the choice of the importance sampling distribution. A good choice is often hard
to achieve especially for high-dimensional integration problems.
Nonparametric estimation of the optimal importance sampling distribution
(known as nonparametric importance sampling) is a reasonable alternative to
parametric approaches. 
In this article nonparametric variants of both the self-normalized and the
unnormalized importance sampling estimator are proposed and investigated.
 A common critique on nonparametric importance sampling is the increased
computational burden compared to parametric methods. We solve this problem to
a large degree by utilizing the linear blend frequency polygon estimator
instead of a kernel estimator. Mean square error convergence properties are
investigated leading to recommendations for the efficient application of
nonparametric importance sampling. Particularly, we show that nonparametric
importance sampling asymptotically attains optimal importance sampling variance.
The efficiency of nonparametric importance sampling algorithms heavily relies on
the computational efficiency of the employed nonparametric estimator. 
The linear blend frequency polygon outperforms kernel estimators
in terms of certain criteria such as efficient sampling and evaluation.
Furthermore, it is compatible with the inversion method for sample generation.
This allows to combine our algorithms with other variance
reduction techniques such as stratified sampling. 
Empirical evidence for the usefulness of the suggested algorithms is obtained
by means of three benchmark integration problems. 
As an application we estimate the distribution of the queue
length of a spam filter queueing system based on real data.

\bigskip


\newpage
\setcounter{page}{1}

\begin{center}
1. INTRODUCTION
\end{center}
Importance Sampling (IS) is a general sampling technique for approximating
expectations
\[
\mathbf{E}_p[\varphi] = I_\varphi = \int \varphi(\mathbf{x}) p(\mathbf{x})
d\mathbf{x}
\]
of some function $\varphi: \mathbb{R}^d \rightarrow \mathbb{R}$ with respect to
a probability density function $p$ on $\mathbb{R}^d$.
It is often applied if direct sampling from distribution $p$ is
computationally too demanding or intractable. But IS is not limited to this
purpose. Unless $\varphi$ is constant, IS can often yield massive
reduction of the estimators variance if applied carefully.
Formally importance sampling is a change of measure. The expectation
$\mathbf{E}_p[\varphi]$ is rewritten as
\[
\mathbf{E}_q[\varphi w] = \int \varphi(\mathbf{x}) w(\mathbf{x}) q(\mathbf{x})
d\mathbf{x}
\]
where $q$ is the probability density function of an importance sampling
distribution (also known as proposal) and
$w(\mathbf{x})=p(\mathbf{x})/q(\mathbf{x})$ the Radon-Nikodym derivative.  
The proposal needs to be chosen so that its support includes the support of
$|\phi| p$ or $p$, which imposes a first constraint on $q$. 
Using importance sampling the integral
$I_\varphi$ can be estimated by
\begin{equation*}
\hat{I}_{\varphi}^{\text{IS}} = \frac1N \sum_{i=1}^{N} \varphi(\mathbf{x}^i)
w(\mathbf{x}^i),
\end{equation*}
where $\{\mathbf{x}^1, \ldots, \mathbf{x}^N\}$ are drawn from proposal $q$.

In Bayesian inference, it is often the case that either $p$ or the
proposal $q$ (or both) are only known up to some constant. In this case an
alternative is the self-normalized importance sampling (SIS) estimator given by
\begin{equation*}
\hat{I}_{\varphi}^{\text{SIS}} = \frac{\sum_{i=1}^{N} \varphi(\mathbf{x}^i)
w(\mathbf{x}^i)}{\sum_{i=1}^{N} w(\mathbf{x}^i)}.
\end{equation*}
The strong law of large numbers implies that both
$\hat{I}_{\varphi}^{\text{IS}}$ and $\hat{I}_{\varphi}^{\text{SIS}}$ converge almost surely to the 
expectation $I_\varphi$ if it is finite. However, this
result is neither of help for assessing the precision of the estimators
for a finite set of samples nor for the rate of convergence. In order to
construct error bounds it is desirable to have a central limit theorem
(CLT) at hand. Under the assumptions that $I_\varphi$ and $\text{Var}_q[\varphi
w]$ are finite a central limit theorem guaranties
$
\sqrt{N} (\hat{I}_{\varphi}^{\text{IS}} - I_{\varphi}) \Rightarrow
\mathcal{N}(0, \sigma_{\text{IS}}^2)
$
where $\sigma_{\text{IS}}^2 = \mathbf{E}_q[\varphi w - I_{\varphi}]^2$
(Rubinstein 1981). The proposal which minimizes the variance $\sigma_{\text{IS}}^2$ is given by
\begin{equation}\label{is:optdis}
q_{\varphi}^{\text{IS}}(\mathbf{x}) = \frac{|\varphi(\mathbf{x})|p(\mathbf{x})}
{\int |\varphi(\mathbf{x})| p(\mathbf{x}) d\mathbf{x}}.
\end{equation}
$q_{\varphi}^{\text{IS}}$ is called the optimal proposal. 
A CLT for the self-normalised IS estimator $\hat{I}_{\varphi}^{\text{SIS}}$ can be
established
$
{\sqrt{N} (\hat{I}_{\varphi}^{\text{SIS}} - I_{\varphi}) \Rightarrow
\mathcal{N}(0, \sigma_{\text{SIS}}^2)}
$
with limiting variance $\sigma_{\text{SIS}}^2 =
\mathbf{E}_q[(\varphi-I_{\varphi}) w]^2$ under the additional assumption that
$\text{Var}_q[w]<\infty$ (Geweke 1989). 
Variance $\sigma_{\text{SIS}}^2$ is minimized by the proposal
\begin{equation}\label{sis:optdis}
q_{\varphi}^{\text{SIS}}(\mathbf{x}) =
\frac{|\varphi(\mathbf{x})-I_{\varphi}|p(\mathbf{x})} {\int
|\varphi(\mathbf{x})-I_{\varphi}|p(\mathbf{x}) d\mathbf{x}},
\end{equation}
provided that the median of $\varphi$ with respect to $p$ exists.
The optimal proposals (\ref{is:optdis}) and (\ref{sis:optdis}) are
merely of conceptual help as the computation of their denominators is
typically at least as difficult as the original integration problem.
Hence, the objective is to find an easy-to-sample density that approximates
the optimal proposals.
Traditionally, a proposal is chosen from some parametric family of
densities $\{q_{\varphi,\theta}, \theta \in \Theta\}$ that satisfy the
assumptions of the central limit theorems or some related conditions. Typically,
it is demanded that the support of $q_{\varphi,\theta}$ includes
the support of $|\varphi|p$ or $|\varphi-I_{\varphi}|p$, respectively, and
that the tails of $q$ do not decay faster than those of $|\varphi|p$. Many
different density classes have been investigated in the literature including
multivariate Student t, mixture, and exponential family distributions (see for
instance Geweke 1989; Stadler and Roy 1993; Oh and Berger 1993). 
The parametrized choice of the proposal can be adaptively revised during the
importance sampling which is known as adaptive IS (Oh and Berger
1992; Kollman et al. 1999). Often expectation $I_\varphi$ needs to
be computed for many different functions $\varphi$ leading to different optimal
proposals. 
As a consequence, it is necessary to investigate the structure of any new
$\varphi$ in order to find a suitable parametric family.

A reasonable alternative that does not rely on prior investigation of the
structure of the integrand is nonparametric importance sampling (NIS).
Nonparametric approximations based on kernel estimators for the construction
of proposals have been used before (West 1992, 1993; Givens and Raftery 1996;
Kim et al. 2000). Under restrictive conditions it has been shown that
nonparametric (unnormalized) IS can not only reduce the variance of the
estimator but may also improve its rate of convergence of the mean square error (MSE) to 
$\mathcal{O}(N^{-(d+8)/(d+4)})$ (Zhang 1996). 
Except for special cases, parametric importance sampling strategies achieve
the standard MC rate of $\mathcal{O}(N^{-1})$, as the optimal proposal is typically not included in the
employed distribution family. There is still a lack of theoretical results for
NIS, particularly for the self-normalized importance sampler. Furthermore,
computationally aspects, that critically effect the performance of NIS, have
only been insufficiently treated in the literature (Zlochin and
Baram 2002).

The competitiveness of NIS compared
to parametric IS heavily relies on the computational efficiency of the
employed nonparametric estimator.
This article introduces NIS algorithms based
on a multivariate frequency polygon estimator which we show to be
computationally superior to kernel estimators. 
Furthermore, our nonparametric estimator allows the combination of NIS
with other variance reduction techniques (such as stratified sampling) which
is another advantage over kernel estimators.  
We investigate NIS not only for IS but also for SIS which has not been done
before. Under loose conditions on the integrand, the MSE convergence properties of the
proposed algorithms are explored. 
The theoretical findings result in distinct suggestions for efficient
application of NIS. 
The large potential of NIS to reduce MC variance is verified empirically 
by means of different integration problems.
Overall, we provide strong evidence that our NIS algorithms solve well-known
problems of existing NIS techniques. This suggests that NIS is a promising
alternative to parametric IS in practical applications.

The remainder of the paper is organized as follows.
In Sections 2 and 3 we propose NIS algorithms for IS
and SIS, respectively, and investigate their MSE convergence properties. In
Section 4 we discuss the applicability of the suggested
algorithms including parameter selection and implementation issues. 
Finally, in Section 5 and 6 we
present simulation results for three toy integration problems and for a spam
filter queueing system based on real data.

\begin{center}
2. NONPARAMETRIC IMPORTANCE SAMPLING
\end{center}
A NIS algorithm based on a kernel density estimator, that approximates the
analytically unavailable optimal proposal $q_{\varphi}^{\text{IS}}$, is
considered in Zhang (1996). Theoretical and empirical evidence of the
usefulness of this approach has been established. In particular, it was proved that NIS
may yield MSE convergence of order  $\mathcal{O}(N^{-(d+8)/(d+4)})$ 
essentially under the very restrictive assumption that $\varphi p$ has compact
support on which $\varphi$ is strictly positive. 
The theoretical results derived in this paper get by with much weaker
assumptions. From a practical point of view a kernel density
estimator is computationally too demanding. For the purpose of NIS it does not 
suffice that the employed nonparametric estimator provides a fast and accurate 
approximation of the distribution of interest. It is also
required to allow efficient sampling as well as fast evaluation at arbitrary 
points. As a computationally more efficient alternative to the kernel
estimator, it is suggested to use a histogram estimator (Zhang 1996). The
drawback of a histogram is its slow convergence rate of $\mathcal{O}(N^{-2/(2+d)})$
compared to kernel estimators, which typically achieve
$\mathcal{O}(N^{-4/(4+d)})$. In this paper we propose the usage of a
multivariate frequency polygon which is known as linear blend frequency
polygon (LBFP) (Terrell 1983 cited in Scott 1992, p. 106). It is constructed
by interpolation of histogram bin mid-points. Being computationally only slightly more expensive
than ordinary histograms, it achieves the same convergence rate as standard
kernel estimators. 
Consider a multivariate histogram estimator with bin height
$\hat{f}^{\text{H}}_{k_1, \ldots, k_d}$ for bin $B_{k_1, \ldots, k_d} =
\prod_{i=1}^d [t_{k_i}-h/2, t_{k_i} + h/2)$ where $h$ is the bin width and
$(t_{k_1}, \ldots, t_{k_d})$ the bin mid-point.
For ${\bf x} \in \prod_{i=1}^{d} [t_{k_i},t_{k_i} + h)$ the LBFP estimator is
defined as
\begin{equation}\label{def:LBFP}
\hat{f} ({\bf x}) = \sum_{j_1, \ldots, j_d \in \{0,1\}} \left[
\prod_{i=1}^{d} \left(\frac{x_i-t_{k_i}}{h}\right)^{j_i}
\left(1-\frac{x_i-t_{k_i}}{h}\right)^{1-j_i} \right]
\hat{f}^{\text{H}}_{k_1+j_1,\ldots,k_d+j_d}.
\end{equation}
It can be shown that $\hat{f}$ integrates to one.

Our NIS algorithm consists of two steps. In the first step the
optimal proposal $q_{\varphi}^{\text{IS}}$ is estimated nonparametrically using samples drawn from a
trial distribution $q_0$ and weighted according to the importance ratio
$q_{\varphi}^{\text{IS}}/q_0$. In the second step an ordinary
importance sampling is carried out, subject to the proposal estimated in
the first step.
Before we can state the algorithm we need to introduce the following
quantities. Let $A_M$ be an increasing sequence of compact sets defined by $A_M
= \{ \mathbf{x} \in \mathbb{R}^{d} : q_{\varphi}^{\text{IS}}(\mathbf{x}) \geq c_M \}$,
where $c_M > 0$ and $c_M \rightarrow 0$ as $M$ goes to infinity. For any function 
$g$ we denote the restriction of $g$ on $A_M$ by $g_M$ and we abbreviate
$q_{M}^{\text{IS}} = q_{\varphi_M}^{\text{IS}}$. Furthermore, the volume of
$A_M$ is denoted by $V_M$. Note that, by definition, $A_M$ converges to the
support of $q_{\varphi}^{\text{IS}}$. The theorems in this section consider the
following algorithm, which is related to the NIS algorithm in Zhang (1996).

\bigskip

\hrule

\smallskip

\noindent
\textbf{Algorithm 1 \-- Nonparametric Importance Sampling}

\noindent
{\em Step 1: Proposal estimation}
	\begin{itemize}
		\item {\bf For $j=1,\ldots,M$}: $\;$ Sample $\tilde{\mathbf{x}}^j \sim q_0$.
		\item Obtain estimate
		$\hat{q}_M^{\text{IS}}(\mathbf{x}) = 
		\frac{\hat{f}_M(\mathbf{x}) + \delta_M}{\overline{\omega}_M + V_M
		\delta_M} \mathbf{1}_{A_M}(\mathbf{x})$,\\
		where $\overline{\omega}_M = 1/M \sum_{j=1}^M \omega_M^j$, $\omega_M^j =
		|\varphi_M(\tilde{\mathbf{x}}^j)| p(\tilde{\mathbf{x}}^j)
		q_0(\tilde{\mathbf{x}}^j)^{-1}$, and
		\begin{eqnarray*}
        \hat{f}_M(\mathbf{x}) &=& \frac{1}{M} \sum_{j_1, \ldots, j_d \in
        \{0,1\}} \left[
\prod_{i=1}^{d} \left(\frac{x_i-t_{k_i}}{h}\right)^{j_i}
\left(1-\frac{x_i-t_{k_i}}{h}\right)^{1-j_i} \right] \\
		&& \qquad \times  \sum_{j=1}^{M} \omega_M^j \mathbf{1}_{\prod_{i=1}^{d}
[t_{k_i},t_{k_i} + h)}(\tilde{\mathbf{x}}^j)
		\end{eqnarray*}
		for ${\bf x} \in \prod_{i=1}^{d} [t_{k_i},t_{k_i} + h)$.
		\end{itemize}
	{\em Step 2: Importance Sampling}
	\begin{itemize}
		\item {\bf For $i=1,\ldots,N-M$}: $\;$ Generate sample $\mathbf{x}^i$ from proposal
		$\hat{q}_M^{\text{IS}}$.
		\item Evaluate 
		$\hat{I}_{\varphi_M}^{\text{NIS}} =
		(N-M)^{-1} \sum_{i=1}^{N-M} \varphi_M(\mathbf{x}^i) p(\mathbf{x}^i)
		\hat{q}_M^{\text{IS}}(\mathbf{x}^i)^{-1}.$
	\end{itemize}
\hrule

\bigskip

Both $A_M$ and $\delta_M$ are required in the proofs of the Theorems below but
they can be omitted in practice.

\bigskip

\noindent
\textbf{Assumption 1}
Both $\varphi$ and $p$ have three continuous and square integrable
derivatives on $\text{supp}(|\varphi| p)$ and $|\varphi| p$ is
bounded. Furthermore, $\int (\nabla^2 |\varphi| p)^4 (|\varphi|p)^{-3}
< \infty$ where $\nabla^2 |\varphi| p = \partial^2 |\varphi| p/ \partial
x_1^2 + \ldots + \partial^2 |\varphi| p/ \partial
x_d^2$.

\noindent
\textbf{Assumption 2} $\mathbf{E} [|\varphi| p q_0^{-1}]^4$ is finite on $\text{supp}(|\varphi| p)$.

\noindent
\textbf{Assumption 3}
As total sample size $N \rightarrow \infty$, bin width $h$ satisfies $h
\rightarrow 0$ and $M h^d \rightarrow \infty$. Additionally, we have $\delta_M > 0$, $V_M \delta_M =
o(h^2)$ and $M^3 (V_M \delta_M)^4 \rightarrow \infty$.

\noindent
\textbf{Assumption 4a}
$c_M$ guaranties $\frac{h^8 + (Mh^d)^{-2}}{\delta_M
c_M^3} = o(\frac{h^4 + (Mh^d)^{-1}}{c_M})$ and $\frac{h^4 + (Mh^d)^{-1}}{c_M}
\rightarrow 0$.

\noindent
\textbf{Assumption 5a}
$c_M$ guaranties $(\int q_{\varphi}^{\text{IS}}
\mathbf{1}_{\{q_{\varphi}^{\text{IS} } < c_M \}})^2 = o(M^{-1} h^4 + (M^2
h^d)^{-1})$.

\medskip

For fixed sample size $M$ and
conditional on the samples $\{ \tilde{\mathbf{x}}^1,
\tilde{\mathbf{x}}^2, \ldots, \tilde{\mathbf{x}}^M \}$ it is not hard to show
 that $\hat{I}_{\varphi_M}^{\text{NIS}}$ is an unbiased estimator with
variance
\begin{equation}\label{nis:var}
\text{Var}[\hat{I}_{\varphi_M}^{\text{NIS}}] = \frac{1}{N-M}
\int \left(\frac{\varphi_M(\mathbf{x}) p(\mathbf{x})}
{\hat{q}_M^{\text{IS}}(\mathbf{x})} - I_{\varphi_M}\right)^2
\hat{q}_M^{\text{IS}}(\mathbf{x}) d\mathbf{x}.
\end{equation}
For the special case $\varphi \geq 0$
we have $q_M^{\text{IS}} = \varphi_M p I_{\varphi_M}^{-1}$ and
(\ref{nis:var}) can be rewritten as
\begin{equation}\label{nis:var2}
\frac{I_{\varphi_M}^2}{N-M}
 \int \frac{(\hat{q}_M^{\text{IS}}(\mathbf{x}) -
 q_M^{\text{IS}}(\mathbf{x}))^2} {\hat{q}_M^{\text{IS}}(\mathbf{x})}
 d\mathbf{x}.
\end{equation}
Under the foregoing assumptions we now prove that the variance
(\ref{nis:var2}) attains convergence rate $\mathcal{O}(N^{-(d+8)/(d+4)})$,
if bin width $h$ is chosen optimally.

\begin{thm}\label{thm:nis}
Suppose Assumptions 1-3, 4a, 5a hold, $\varphi \geq 0$, and
$q=q_{\varphi}^{\text{IS}}$. We obtain
\begin{eqnarray*}
\mathbf{E}[\hat{I}_{\varphi_M}^{\text{NIS}} - I_{\varphi}]^2 &=&
\frac{I_{\varphi}^2} {N-M} \left\{h^4 H_1 + \frac{2^d}{3^d M h^d} H_2 \right\} \times (o(1) + 1)
\end{eqnarray*}
and the optimal bin width
\[ h^{*} = \left(\frac{d H_2 2^d }{4 H_1 3^d}\right)^{\frac{1}{d+4}}
M^{-\frac{1}{d+4}} \]
where 
\[H_1=\frac{49}{2880} \sum_{i=1}^{d} \int \frac{(\partial_i^2
q)^2}{q}+ \frac{1}{64} \sum_{i \not= j} \int \frac{\partial_i^2
q \partial_j^2 q}{q}, \; H_2= \int
\frac{q}{q_0}.\]
\end{thm}
\proof See Appendix A.

\noindent
A direct implication of Theorem \ref{thm:nis} is the following
corollary.

\begin{cor}\label{cor:nis}
Under the assumptions of Theorem \ref{thm:nis} and the further assumption
that $M/N \rightarrow \lambda$ $(0 < \lambda < 1)$,
and $h=h^{*}$ we yield 
\[\lim_{N\rightarrow \infty} N^{\frac{d+8}{d+4}}
\mathbf{E}\left[\hat{I}_{\varphi_M}^{\text{NIS}} - I_{\varphi}
\right]^2 = \lambda^{-\frac{4}{d+4}} (1-\lambda)^{-1} \times I_{\varphi}^2 D\]
and optimal proportion $\lambda^{*} = 4/(d+8)$,\\
where
$ D = \left\{ (d/4)^{4/(d+4)} + (d/4)^{-d/(d+4)} \right\} \left[ H_1^d (2^d
3^{-d} H_2)^4 \right]^{1/(d+4)}$.
\end{cor}

\noindent
We remark that under much stronger assumptions corresponding results for NIS
based on kernel estimators were obtained in Zhang (1996).

We now move to a more general case. Assume $\varphi \geq 0$ (and $\varphi \leq
0$) does not hold. For this case we show that the NIS
algorithm asymptotically achieves the minimum importance sampling variance.  By
substituting the optimal IS distribution $q_{\varphi}^{\text{IS}}$ into 
variance $\sigma^2_{\text{IS}}$ and writing shorthand  $\overline{I}_{\varphi}
= \int |\varphi(\mathbf{x})| p(\mathbf{x}) d\mathbf{x}$, we see the optimal
variance of the IS estimator to be $\overline{I}_{\varphi}^2 -
I_{\varphi}^2$.  

\medskip 
\noindent
\textbf{Assumption 4b}
$c_M$ guaranties $\frac{h^8 + (Mh^d)^{-2}}{\delta_M
c_M^5} = o(\frac{h^4 + (Mh^d)^{-1}}{c_M^3})$ and $\frac{h^4 +
(Mh^d)^{-1}}{c_M^3} \rightarrow 0$.

\noindent
\textbf{Assumption 5b}
$c_M$ guaranties $(\int q_{\varphi}^{\text{IS}}
\mathbf{1}_{\{q_{\varphi}^{\text{IS}} < c_M \}})^2 = o(M^{-1} h^2 + (M^2
h^d)^{-1})$.

\begin{thm}\label{thm:nis2}
Suppose that Assumptions 1-3, 4b, 5b hold, $\varphi$ does not have a
definite sign, and $q = q_{\varphi}^{\text{IS}}$. Then we obtain
\begin{eqnarray*}
\mathbf{E}[\hat{I}_{\varphi_M}^{\text{NIS}} - I_{\varphi}]^2 &=&
\frac{1}{N-M} \left[ (\overline{I}_{\varphi}^2 - I_{\varphi}^2) \ + \
I_{\varphi}^2 \left\{ h^2 \overline{H}_1 + \frac{2^d}{3^d M h^d}
\overline{H}_2 \right\} \times (1 + o(1)) \right]
\end{eqnarray*}
and the optimal bin width
\[ h^{**} = \left(\frac{d \overline{H}_2 2^{d-1}}{\overline{H}_1
3^d }\right)^{\frac{1}{d+2}} M^{-\frac{1}{d+2}} \]
where $\overline{H}_1 = -\left(\int f_{\varphi}^2 \frac{\nabla^2 q}{8 q^2} +
\int f_{\varphi} \frac{\nabla^2 q}{4 q} \right)$, 
$\overline{H}_2 = \left( \int \frac{q}{q_0} - 2 \int \frac{f_{\varphi}}{q_0} -
\int \frac{f^2_{\varphi}}{q_0 q} \right)$, and 
${f_{\varphi} =
\left(\frac{\varphi p}{I_{\varphi}} - \frac{|\varphi|
p}{\overline{I}_{\varphi}}\right)}$.
\end{thm}
\proof See Appendix A.

As a consequence of Theorem \ref{thm:nis2}, the NIS algorithm does not lead to a
MSE rate improvement for functions $\varphi$ which take positive and negative
values. But if the optimal bin width $h^{**}$ is used, we have 
\begin{equation*}
\mathbf{E}[\hat{I}_{\varphi_M}^{\text{NIS}} - I_{\varphi}]^2 = 
\frac{\overline{I}_{\varphi}^2 - I_{\varphi}^2}{N-M}  + o(N^{-1}).
\end{equation*}
That is, the optimal IS variance is achieved asymptotically.
Unlike Theorem \ref{thm:nis}, the optimal proportion $\lambda$
cannot be computed analytically due to its dependency on $N$. But theoretically
it can be computed as 
$\lambda^{**} = \text{argmin}_{\lambda} G(N,h^{**}, \lambda)$
where $G=\mathbf{E}[\hat{I}_{\varphi_M}^{\text{NIS}} - I_{\varphi}]^2$.
Clearly, $\lambda^{**}$ decreases in $N$. Note, that for 
the optimal asymptotic variance to be achieved, it suffices that $0 < \lambda <
1$.

Corollary \ref{cor:nis} and Theorem \ref{thm:nis2} suggest that IS based Monte
Carlo integration can be much more efficient for functions $\varphi \geq 0$
(and  $\varphi \leq 0$) than for arbitrary functions. 
This stems from the fact that for non-negative (non-positive) 
functions the usage of the optimal proposal leads to a zero variance estimator. 
By approximating the optimal
proposal with a consistent estimator it is therefore not surprising that the
standard MC rate can be surmounted. Consequently, it should be reasonable to decompose
$\varphi$ into positive and negative part, $\varphi = \varphi^{+} -
\varphi^{-}$, and apply {Algorithm 1} to $\varphi^{+}$ and $\varphi^{-}$
separately. Since then, we can expect to achieve the superior rate
$\mathcal{O}(N^{-(d+8)/(d+4)})$. 
Note that the partitioning of $\varphi$ needs not to be done analytically. 
It may be carried out implicitly in Step 1 of the algorithm.
This approach, denoted by NIS+/-, is
investigated in a simulation study in Section 5.

\begin{center}
3. NONPARAMETRIC SELF-NORMALIZED IMPORTANCE SAMPLING
\end{center}
Many problems in Bayesian inference can be written as the expectation of some 
function of interest, $\varphi$, with respect to the posterior distribution $p$
which is only known up to some constant. This leads to the evaluation of
integrals
\[ \mathbf{E}_p[\varphi] = \frac{\int \varphi(\mathbf{x})
\tilde{p}(\mathbf{x}) d \mathbf{x}} {\int \tilde{p}(\mathbf{x}) d\mathbf{x}},
\]
where $\tilde{p} = \alpha p$ with unknown constant $\alpha$.
Self-normalized IS is a standard approach for solving such problems. It is
often suggested to choose the proposal close to the posterior. But from the
CLT we know that one can do better by
choosing it close to the optimal proposal which is
proportional to $|\varphi-I_{\varphi}| p$. 
Next, we introduce a nonparametric self-normalized IS (NSIS) algorithm.

In analogy to the definition of $A_M$ we define 
$\widetilde{A}_M =  \{ \mathbf{x}  \in \mathbb{R}^{d} :
q_{\varphi}^{\text{SIS}}(\mathbf{x}) \geq \tilde{c}_M \}$,  where $\tilde{c}_M
> 0$ and $\tilde{c}_M \rightarrow 0$ as $M$ goes to infinity. 
Its volume is denoted by $\widetilde{V}_M$.

\bigskip

\hrule

\smallskip

\noindent
\textbf{Algorithm 2 \-- Nonparametric Self-Normalized Importance Sampling}

\noindent
{\em Step 1: Proposal estimation}
	\begin{itemize}
		\item {\bf For $j=1,\ldots,M$}: $\;$ Sample $\tilde{\mathbf{x}}^j \sim q_0$.
		\item Obtain estimate 
		$ \hat{q}_M^{\text{SIS}}(\mathbf{x}) = 
		\frac{\hat{f}_M(\mathbf{x}) + \delta_M}{\overline{\omega}_M + \widetilde{V}_M
		\delta_M} \mathbf{1}_{\widetilde{A}_M}(\mathbf{x})$,\\
		where $\overline{\omega}_M = 1/M \sum_{j=1}^M \widetilde{\omega}_M^j$,
		$\widetilde{\omega}_M^j = |\varphi_M(\tilde{\mathbf{x}}^j) - 
		\breve{I}_{\varphi_M}| \tilde{p}(\tilde{\mathbf{x}}^j)
		q_0(\tilde{\mathbf{x}}^j)^{-1}$, $\hat{f}_M(\mathbf{x})$ analogous to
		Algorithm 1, and 
		\[\breve{I}_{\varphi_M} = \frac{ \sum_{j=1}^{M}
		\varphi_M(\tilde{\mathbf{x}}^j) \tilde{p}(\tilde{\mathbf{x}}^j)
		q_0(\tilde{\mathbf{x}}^j)^{-1}} {\sum_{j=1}^{M}
		\tilde{p}(\tilde{\mathbf{x}}^j) q_0(\tilde{\mathbf{x}}^j)^{-1}}.\]
		\end{itemize}
	{\em Step 2: Self-Normalized Importance Sampling}
	\begin{itemize}
		\item {\bf For $i=1,\ldots,N-M$}: $\;$ Generate sample $\mathbf{x}^i$ from proposal
		$\hat{q}_M^{\text{SIS}}$.
		\item Evaluate 
 		\[ \hat{I}_{\varphi_M}^{\text{NSIS}} =
 		\frac{ \sum_{i=1}^{N-M} \varphi_M(\mathbf{x}^i) \widetilde{w}_M(\mathbf{x}^i)}
 		{\sum_{i=1}^{N-M}  \widetilde{w}_M(\mathbf{x}^i)}\] 
		where $\widetilde{w}_M(\mathbf{x}^i) = \tilde{p}(\mathbf{x}^i)
		\hat{q}_M^{\text{SIS}}(\mathbf{x}^i)^{-1}$.
	\end{itemize}
\hrule

\bigskip

Both the SIS and NSIS estimator produce biased estimates. But, however, the
estimators are asymptotically unbiased. 
Under Assumptions 1-3 (with $p$, $|\varphi|$, $c_M$, $V_M$
replaced by $\tilde{p}$, $|\varphi-I_{\varphi}|$, $\tilde{c}_M$,
$\widetilde{V}_M$) it is easy to verify that, conditional on the samples
$\{\tilde{\mathbf{x}}^1, \tilde{\mathbf{x}}^2,  \ldots,
\tilde{\mathbf{x}}^M\}$, the CLT of Geweke (1989) holds for
$\hat{I}_{\varphi_M}^{\text{NSIS}}$. The asymptotic variance of the CLT can be
written as
\begin{equation}\label{nnis:variance}
\sigma_{\text{SIS}}^2 = \tilde{I}_{\varphi_M}^2 \left[ 1 + \int
\frac{(q_M^{\text{SIS}}(\mathbf{x}) -
\hat{q}_M^{\text{SIS}}(\mathbf{x}))^2} {\hat{q}_M^{\text{SIS}}(\mathbf{x})}
d\mathbf{x} \right]
\end{equation}
with $\tilde{I}_{\varphi_M} = \int |\varphi_M(\mathbf{x}) -
I_{\varphi_M}|p(\mathbf{x}) d \mathbf{x}$ the median of $\varphi$.
Consequently, $\tilde{I}_{\varphi_M}^2$ is the (asymptotically) optimal variance
that can be achieved by self-normalized importance sampling. 
Unless $\varphi$ is constant, it is impossible to build up a zero
variance estimator based on SIS. This renders it unnecessary to investigate
separately the MSE convergence of NSIS for non-negative and arbitrary
functions.

The structure of
$\sigma^2_{\text{SIS}}$ is very similar to the structure of the variance in
(\ref{nis:var2}) but the weights $\widetilde{\omega}_M^j$ introduce
inter-sample  dependencies which make the reasoning in the proofs of Theorem
\ref{thm:nis} and Theorem \ref{thm:nis2} not directly applicable. 
However, similarly to Theorem \ref{thm:nis2} we can show that the NSIS
asymptotically attains optimal variance for certain bin width $h$ and 
proportion $0<\lambda<1$.

\begin{thm}\label{thm:nsis}
Suppose that Assumptions 1-3, 4a, 5a (with $p$, $|\varphi|$, $c_M$, $V_M$
replaced by $\tilde{p}$, $|\varphi-I_{\varphi}|$, $\tilde{c}_M$,
$\widetilde{V}_M$) hold, and $q = q_{\varphi}^{\text{SIS}}$. Then we obtain
\begin{eqnarray*}
\mathbf{E}[\hat{I}_{\varphi_M}^{\text{NSIS}} - I_{\varphi}]^2 &=&
\frac{\tilde{I}_{\varphi}^2}{N-M} \left[ 1 \ + \
 h^4 H_1 + \frac{2^d}{3^d M h^d}
H_2  \right] \times (1 + o(1))
\end{eqnarray*}
and the optimal bin width
\[ \tilde{h}^{*} = \left(\frac{d H_2 2^d }{4 H_1 3^d}\right)^{\frac{1}{d+4}}
M^{-\frac{1}{d+4}} \]
where $H_1$ and $H_2$ are defined as in Theorem \ref{thm:nis} (with
$q_{\varphi}^{\text{IS}}$ replaced by $q_{\varphi}^{\text{SIS}}$).
\end{thm}
\proof See Appendix A.

First, note that analogous to Theorem \ref{thm:nis2}, there is no analytic
solution for the optimal $\lambda$. 
Second, the theorem implies that with NSIS the MSE rate cannot be improved.
Therefore, NSIS is (at least asymptotically) less efficient than NIS+/-.
There is consequently no reason to apply NSIS in cases where NIS+/-
is applicable. 
However, this does not impair the usefulness of NSIS in cases where 
normalization is required due to unknown constants.

\begin{center}
4. APPLYING NONPARAMETRIC IMPORTANCE SAMPLING
\end{center}
In this section we discuss what is required for implementing NIS/NSIS.
First, one need to take care of the selection of $q_0$, $h$, and $\lambda$.
Second, an implementation of the LBFP estimator which allows the
generation of samples is required. Given these ingredients the implementation of
Algorithm 1 and 2 is straightforward.

\begin{center}
4.1 Parameter Selection
\end{center}

\noindent
(i) From a practical point of view trial distribution $q_0$ should be
chosen such that its support is close to the support of
$|\varphi|p$ or $|\varphi-I_{\varphi}|p$, respectively, and such that it has
heavier tails than the corresponding optimal proposal. But it is not required
that $q_0$ emulates any structure of the optimal proposal. Obviously, the
choice should also comply with Assumption 2. Note that the
expectations in the assumptions may not exist, if $q_0$ is too close
to the optimal proposals.
In addition, it is important to choose an easy-to-sample density.
For low-dimensional problems, even a uniform distribution may suffice.

\noindent
(ii) As the optimal bin width incorporates unknown quantities dependent on
the optimal proposal, it typically cannot be computed analytically. The unknown
quantities can be estimated using the plug-in method based on the samples of
{Step 1} of the algorithms, as suggested in Zhang (1996). If the second
derivative of the optimal proposal is unknown, the plug-in method cannot be applied. In
this case, a Gaussian reference rule is an alternative.

\noindent
(iii) Except for the case investigated in Theorem \ref{thm:nis} and Corollary
\ref{cor:nis}, where the optimal proportion $\lambda^{*}$ is given by a
beautifully easy expression only depending on the problem dimension, it is not
clear how to choose $\lambda$. However, from the MSE
error expressions in the theorems we know that $\lambda^{*}$ (from Corollary
\ref{cor:nis}) serves as an upper bound. Empirical evidence suggests that
$\lambda$ should never exceed $.25$.

\noindent
(iv) In practical applications the restriction of the estimator on a
compact set $A_M$ can be omitted as the induced bias can be made arbitrarily
small and particularly smaller than the desired precision of the integral value.
Hence, the sequence $c_M$ does not need to be defined. 
Sequence $\delta_M$ can also be skipped in practice as mentioned before.

\begin{center}
4.2 Implementing the LBFP estimator
\end{center}
The implementation of the LBFP estimator
$\hat{f}$ should take into account efficient sampling and
evaluation. Given the multivariate histogram defined through
bin heights $\hat{f}^{\text{H}}_{k_1, \ldots, k_d}$ the implementation
of the evaluation of $\hat{f}$ is simple (see (\ref{def:LBFP})). 
We emphasize that for storing $\hat{f}$ on a computer it suffices to store the
underlying histogram. Sampling from a LBFP is more involved than evaluation and to
the author's knowledge this has not been discussed in the literature until now.
We propose to apply the inversion method. 
The crucial fact is that a LBFP can be written as a product of
(conditional) univariate frequency polygons (FP)
\begin{equation*}
\hat{f}(\mathbf{x}) = \hat{f}^{\text{FP}}(x_1) \prod_{i=2}^{d}
\hat{f}^{\text{FP}}(x_i | x_{1:i-1})
\end{equation*}
with $\{x_{1:i-1}\} = \{x_1, \ldots, x_{i-1}\}$.
This representation suggests to produce draws from $\hat{f}$ by sampling
recursively from the univariate FPs $\hat{f}^{\text{FP}}$ using their inverse
cumulative distribution functions.
A FP is a convenient object as it is just a linear interpolated univariate 
histogram. Furthermore we have
\begin{equation}\label{relation:uniFP-LBFP}
\hat{f}^{\text{FP}}(x_i|x_{1:i-1}) = \frac{\hat{f}(x_{1:i})}
{\hat{f}(x_{1:i-1})}
\end{equation}
where $\hat{f}(x_{1:i})$ are (marginalized) LBFPs, $i=1, \ldots, d$.
We will see below that the $\hat{f}^{\text{FP}}(x_i | x_{1:i-1})$ are not
required itself but their cumulative distribution functions $\hat{F}(x_i |
x_{1:i-1})$. As FPs are piecewise linear functions and due to relation
(\ref{relation:uniFP-LBFP}) the latter are obtained without difficulty provided
that LBFPs $\hat{f}(x_{1:i})$  can be evaluated. Hence it is required to
calculate the marginalized histograms underlying the LBFPs $\hat{f}(x_{1:i})$.
These are specified through bins $B_{k_1, \ldots, k_i}$ and bin heights 
$\hat{f}^{\text{H}}_{k_1, \ldots, k_i}$.

Let $\mathbf{y} = \{y_1, \ldots, y_d \} \in [0,1)^d$ and $y_i \in [\hat{F}(t_{k_i} |
x_{1:i-1}), \hat{F}(t_{k_i+1} | x_{1:i-1}))$.
We now describe how the inverse cumulative distribution functions
$\hat{F}^{-1} (\cdot| x_{1:i-1})$ of $\hat{f}^{\text{FP}}(x_i | x_{1:i-1})$ can
be evaluated at $y_i$ by making use of $\hat{F}(x_i |
x_{1:i-1})$.
It is easy to see that, for $x_i \in [t_{k_i}, t_{k_i+1})$,
$\hat{f}^{\text{FP}}(x_i | x_{1:i-1})$ is a linear function with intercept
$\alpha$ and slope $\beta$ where
\begin{equation*}
\alpha = \frac{\hat{f}(x_{1:i-1}, t_{k_i})} {\hat{f}(x_{1:i-1})} \qquad
\text{and} \qquad \beta = \frac1h \left[\frac{\hat{f}(x_{1:i-1}, t_{k_i+1})}
{\hat{f}(x_{1:i-1})} -\alpha \right].
\end{equation*}
Hence $\hat{F}^{-1} (y_i | x_{1:i-1})$ is the solution of the
quadratic equation
\begin{equation}\label{quadratequ} 
y_i - \hat{F}(t_{k_i} | x_{1:i-1}) = \int_{t_{k_i}}^{z} \hat{f}^{\text{FP}}(x_i
| x_{1:i-1}) d x_i = \alpha z + \frac{\beta}{2} z^2
\end{equation}
which is given by
\begin{equation}\label{invcum}
\hat{F}^{-1}(y_i | x_{1:i-1}) = 
\begin{cases}
	-\frac{\alpha}{\beta} + \text{sgn}(\beta ) \sqrt{\frac{\alpha^2}{\beta^2}-2
	\frac{\gamma_1-y_i}{\beta}} & \quad \text{for} \quad \beta \not= 0, \\ 
	 \left[(\gamma_2-y_i)t_{k_i} + (y_i-\gamma_1)t_{k_{i}+1}
	 \right] / (\gamma_2-\gamma_1) & \quad \text{for} \quad \beta=0,
\end{cases}
\end{equation}
where $\gamma_1 = \hat{F}(t_{k_i} | x_{1:i-1})$ and 
$\gamma_2=\hat{F}(t_{k_{i}+1} | x_{1:i-1})$.

Summarizing, a sample $\mathbf{x}^j$ from the LBFP $\hat{f}$ is obtained
 through the following recursion. Let $\mathbf{y}^j$ be a sample from the
 uniform distribution on $[0,1)^d$. Then, for $i=1, \ldots, d$:
\begin{enumerate}
  \item Compute the marginalized histogram associated with LBFP $\hat{f}(x_{1:i})$.
  \item Calculate cumulative distribution function $\hat{F}(x_i |
  x_{1:i-1}^j)$ (or $\hat{F}(x_1)$ for $i=1$) at the (marginal) bin mid points
  $t_{k_i}$ using (\ref{relation:uniFP-LBFP}).
  \item Evaluate $x_i^j = \hat{F}^{-1}(y_i^j|x_{1:i-1}^j)$ (or $x_1^j =
  \hat{F}^{-1}(y_1^j)$ for $i=1$) using (\ref{invcum}).
\end{enumerate}
We remark that for generating $N$ samples Step 1 needs only
to be carried out once as it is independent of the particular sample
$\mathbf{x}^j$. Our \texttt{C++} implementation of the LBFP is available on request.

\begin{center}
4.3 Computational Remarks
\end{center}

Now the computational complexity of the LBFP is discussed. 
For $h=h^{*}$ it can be shown that the complexity
for generating $N$ samples from a LBFP is of order $\mathcal{O}(2^d d^2
N^{(d+5)/(d+4)})$ (see Appendix B for details). 
The complexity of evaluation is of lower order.
Compared to crude MC which has
$\mathcal{O}(dN)$ sampling from a LBFP is only slightly more expensive for $d$
small. For kernel estimators sampling and
evaluation is of order $\mathcal{O}(dN^2)$ (Zlochin and Baram 2002) proving
that the LBFP is computationally more efficient for all relevant $d$ and $N$. Note, more
efficient sampling from kernel estimates is possible using regularization with
whitening (see for instance Musso et al.\ 2001). But this can induce severe
bias if the target distribution is non-Gaussian.

Besides asymptotic complexity properties there are other computational aspects
which are of relevance in practice. On computer systems, the evaluation of
functions such as \texttt{exp} and \texttt{pow} is known to be much more
expensive than standard arithmetic operations. 
Contrary to most parametric IS approaches, nonparametric IS methods do not
require calls to those functions.

\begin{center}
5. SIMULATIONS
\end{center}
We consider three toy examples in order to test our nonparametric
procedures against (parametric) alternatives. The first two examples are
designed to evaluate certain properties of the NIS algorithm and to demonstrate
the degraded performance of the NSIS algorithm. 
The third example is a two-dimensional benchmark problem for self-normalized
importance sampling.

A reasonable measure for the effect of a variance reduction technique is
the relative efficiency (RE). It is defined as the ratio of the crude MC MSE to
the MSE of the method of interest. In the case that both estimators are
unbiased, the RE is also known as variance reduction factor. 
The performance of the different algorithms will be measured by RE and
computation time. 
In all examples the simulation is done for sample sizes 
{$N=$ 1,000}, {$N=$ 5,000}, and {$N=$ 10,000}.
All computation were carried out on a Dell Precision PWS390, Intel CPU
2.66GHz, and the algorithms are coded in \texttt{C++}. For pseudo random number
generation we used the Mersenne Twister 19937 (Matsumoto and
Nishimura 1998). All computation times are reported in milliseconds.

\noindent
\textbf{Example 1.} As our first example we consider a simple integrand that is
to be integrated with respect to the standard normal distribution of dimension $d$. 
The integrand is defined by $\varphi(\mathbf{x}) = x_1
\mathbf{1}_{[-1,1]^d}(\mathbf{x})$.
It takes positive and negative values on the $d$-dimensional unit cube.
This allows the evaluation of the strategy to apply Algorithm 1 separately to
the positive and negative part of the integrand (NIS+/-).
In our simulation $d$ varies from $1$ to $8$. The trial distribution $q_0$ is
set to the uniform distribution on $[-1,1]^d$  and the bin width $h$ is chosen
with the plug-in method. $\lambda$ is set to .15  and to the optimal value
$4/9$ for NIS and NIS+/-, respectively. In order to obtain comparable results,
for NIS+/- the total sample size is equally spread to the integration of the 
positive and negative part.

\begin{figure}[h!]
\centering
\includegraphics[
     height=180pt, width=360pt, keepaspectratio]
     {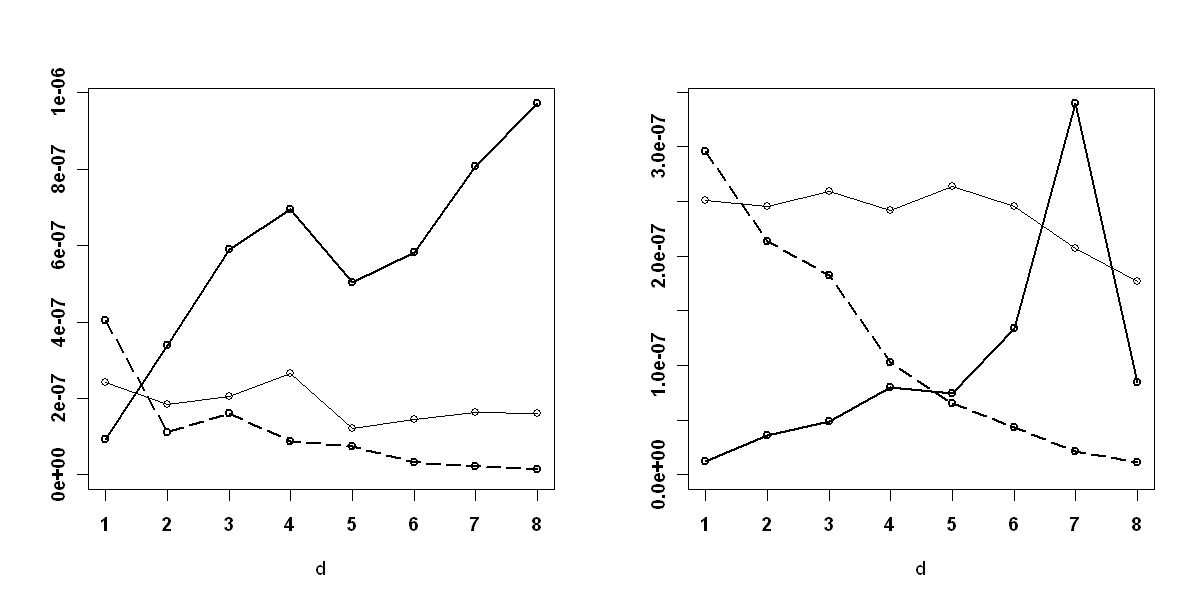}
\caption{\footnotesize Computational efficiency (measured by MSE $\times$
computation time) of MC (solid line), IS (dashed line), and NIS+/- (heavy
line) for $N=$ 1,000 (left), and $N=$ 10,000 (right) for Example 1. All figures
are computed/averaged over 100 independent runs.}
\label{fig:PosNegComEff}
\includegraphics[
     height=180pt, width=360pt, keepaspectratio]
     {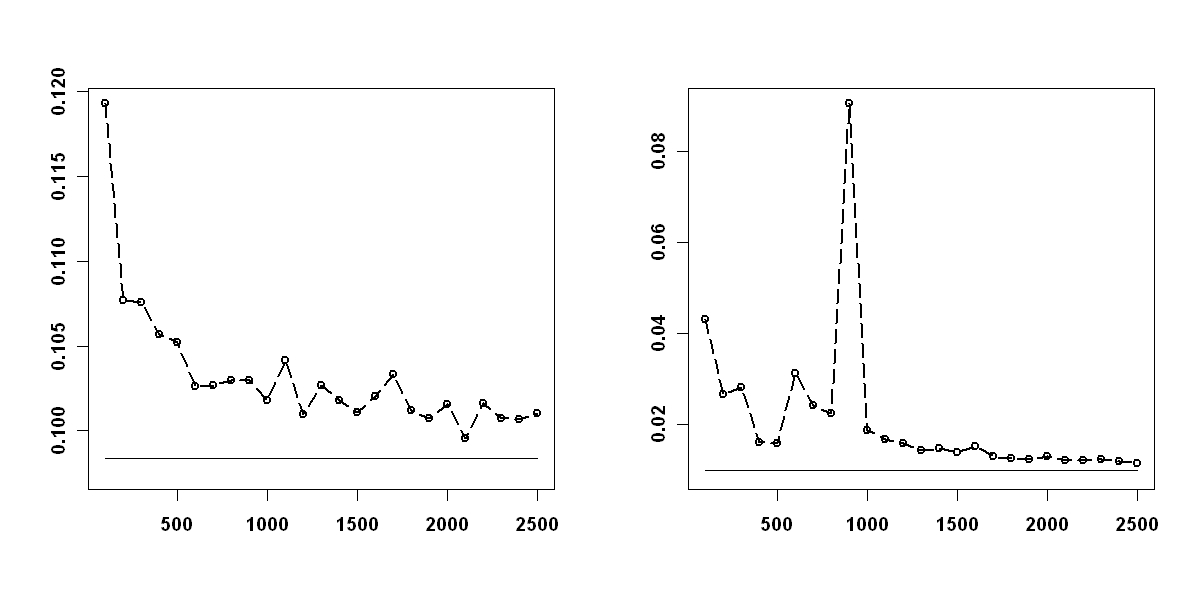}
\caption{\footnotesize Convergence of NIS variance towards optimal IS variance
for $d=1$ (left) and $d=4$ (right) for Example 1. All figures are computed over
10,000 independent runs.}
\label{fig:PosNegOptVarConv}
\end{figure}

Table \ref{table:posneg} shows the RE and computation times for MC, NIS,
NIS+/-, and ordinary IS (subject to the uniform distribution on $[-1,1]^d$).
The RE figures for NIS+/- report large variance reduction which is
present at least up to dimension $d=8$. Even if we take computation time into account,
we find significant efficiency improvement: 
For instance, for $d=4$ and {$N=$ 10,000} we obtain RE of $22$ whereas the
computation time surplus factor is about $7$. Also note, that IS becomes more
favorable as $d$ increases. 
In order to investigate the computationally efficiency we plotted MSE $\times$
computation time (Figure \ref{fig:PosNegComEff}). Contrary to RE, 
smaller values are favourable. We observe that the critical dimension, up to
which NIS+/- is computationally more efficient than the other methods, strongly
depends on the magnitude of $N$. Whereas for {$N=$ 1,000} one would prefer
NIS+/- to IS only for $d=1$, for {$N=$ 10,000} one would do so up to $d=4$.
Finally, the convergence of the NIS variance towards the 
optimal IS variance is examined. The minimum IS variance 
$\overline{I}_{\varphi}^2 - I_{\varphi}^2$ is approximately .098 and 0.0099
for $d=1,4$, respectively. In Figure \ref{fig:PosNegOptVarConv}  the estimated
variances of NIS$\times (1-\lambda)N$ are plotted for 100 $\leq N \leq$ 2,500.
The plots indicate rapid convergence to the optimal values. 
For comparison: the variance of crude MC$\times N$ is roughly .198 (for
$d=1$) and .063 (for $d=4$).

\noindent
\textbf{Example 2.} This example is concerned with the pricing of a call option
within the Black-Scholes model.
Given interest rate $r$ and volatility $\sigma$ the
evolution of a stock is described by the stochastic differential equation (SDE)
$dS(t) / S(t) = r dt + \sigma dW(t)$ with standard Brownian motion
$W$. The solution of the SDE is given by
$
S(T) = S(0) \exp[(r-0.5 \sigma^2)T + \sigma \sqrt{T} Z]
$
where $Z$ is a standard normal random variable.
At time $T$, the call option pays the amount $(S(T)-K)^{+}$ depending
on the strike level $K$. 
The price of the option at time $0$ is given by the
expectation $\mathbf{E}[F(Z)]$ of the discounted payoff $F(Z)=\exp(-rT)
(S(T)-K)^{+}$.
That is, the
pricing problem reduces to the integration of a payoff function with respect to
the standard normal distribution.
Parametric IS is a standard variance reduction technique for
option pricing. A shifted standard normal
distribution is often used as proposal. This approach is known as
change of drift technique.
In our simple model the (asymptotically) optimal drift is given by 
$\text{argmax}_z \log F(z) - .5 z^2$ (Glasserman et al. 1999). 
We state the simulation results for the optimal change of
drift IS (CDIS) as parametric benchmark.

For our simulation we set $S(0)=100$, $r=.1$, $\sigma = .2$, $T=1$. The 
option price is estimated for the strikes $K_1 = 90$ and $K_2 = 130$. 
For $K_1$ the option is said to be in the money ($K_1 < S(0)$) where for $K_2$
it is called out-of-the money ($K_2 > S(0)$). The latter case is particularly
suited for IS techniques, as crude MC fails to satisfactorily sample into
the domain that affects the option price.
$q_0$ is set to the uniform distribution on $[-5,5]$ and bin width $h$ is
selected using the plug-in method. $\lambda$  is set to the optimal value $4/9$
for the NIS and to $.05$ for NSIS.

\begin{figure}
\centering
\includegraphics[
     height=180pt, width=360pt, keepaspectratio]
     {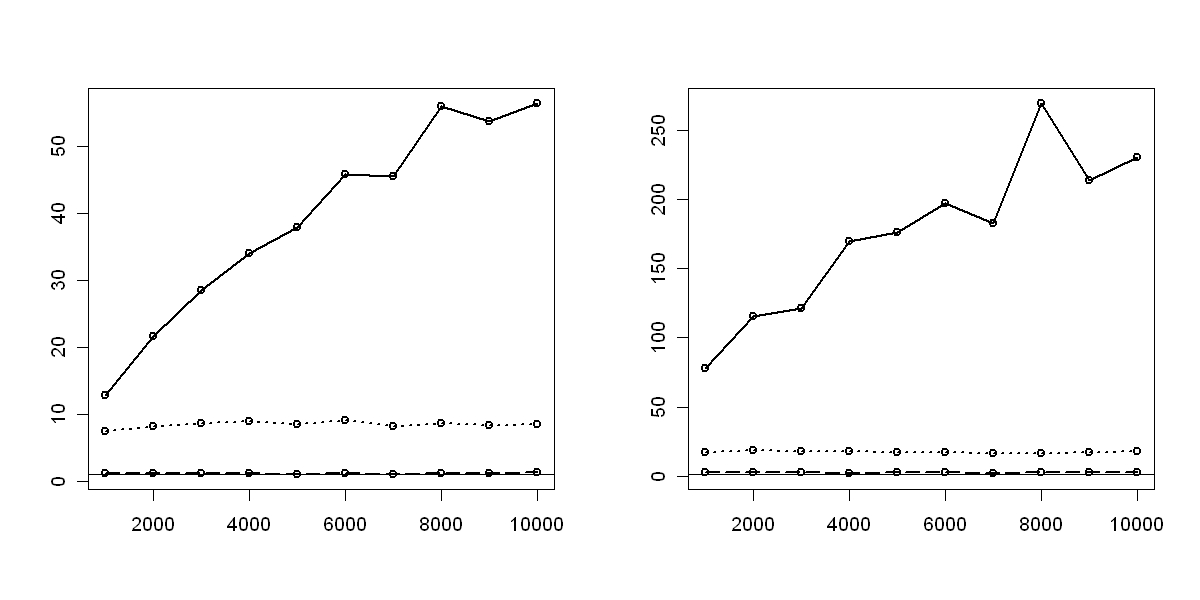}
\caption{\footnotesize Relative efficiency of CDIS (dotted line), NIS (heavy
line), NSIS (dashed line), and crude MC (solid line)  for Example 2 (strike
$K_1$ (left), strike $K_2$ (right)) and 1,000 $ \leq N \leq$ 10,000. All
figures are computed over 1,000 independent runs.}
\label{fig:OptionPricingVarRedPlot90}

\includegraphics[
     height=180pt, width=270pt, keepaspectratio]
     {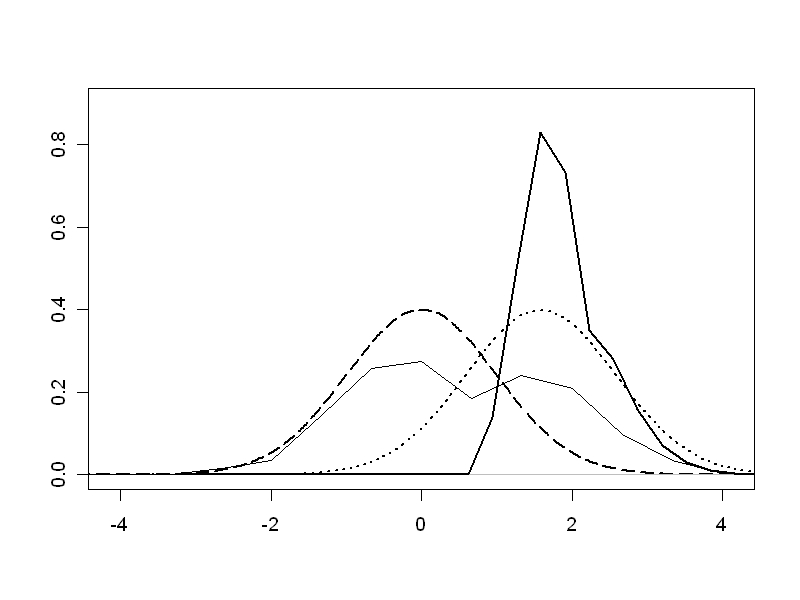}
\caption{\footnotesize Standard normal distribution (dashed line), optimally
shifted normal distribution (dotted line), linear blend frequency
polygon estimates ($N=$ 5,000) of the optimal proposals
$q_{\varphi}^{\text{SIS}}$ (solid line), and $q_{\varphi}^{\text{IS}}$ (heavy
line) for Example 2.}
\label{fig:ProposalPlot}

\centering
\includegraphics[
     height=180pt, width=270pt, keepaspectratio]
     {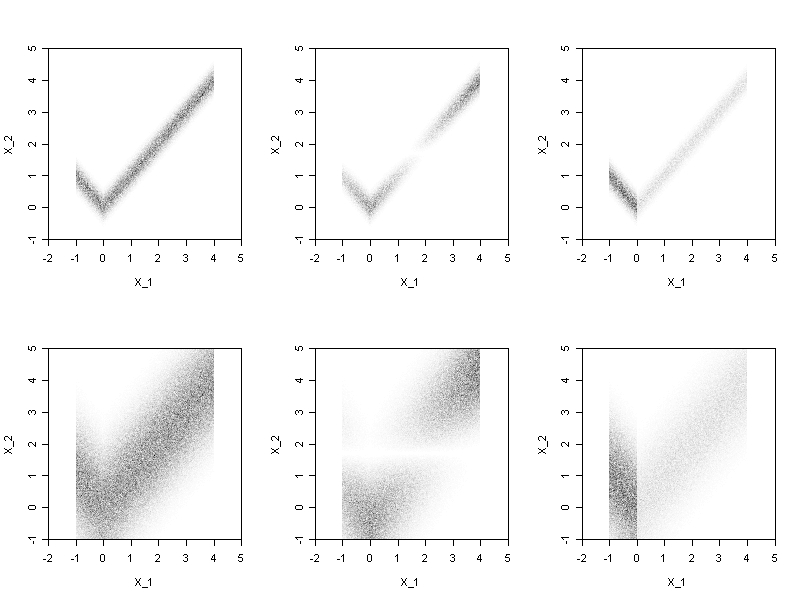}
\caption{\footnotesize Example 3: The upper plots are for the case $a=.75$ and
the lower plots for $a=3.5$. From left to right we have density $p(x_1, x_2)$
and the optimal proposals $q_{\varphi_1}^{\text{SIS}}$ and 
$q_{\varphi_2}^{\text{SIS}}$.}
\label{fig:bayesian}

\end{figure}

The efficiency improvements of the IS methods relative to crude
Monte Carlo integration (RE) are shown in Figure
\ref{fig:OptionPricingVarRedPlot90}. 
Whereas parametric IS methods and NSIS yield constant reduction factors,
NIS realizes increasing relative efficiency which coincides with its theoretical 
superior convergence rate. Particularly for the out-of-the money scenario, NIS
achieves massive variance reduction. Establishing only slight variance reduction
NSIS is worst. This confirms our recommendation to avoid NSIS where NIS is
applicable.
Figure \ref{fig:ProposalPlot} shows the proposals employed 
in the simulation for strike $K_2$. The optimal IS proposal is single-moded 
and can be reasonably approximated by some Gaussian distribution. This
explains the satisfying performance of IS methods based on Gaussian proposals
reported in the literature. However, NIS significantly outperforms CDIS.
For more complex payoffs implying multimodal optimal proposals, the
advantage of NIS should be even more pronounced.
Computation times for different sample sizes are reported in Table
\ref{table:optionPricingTimes}. First, notice that CDIS is much more expensive
than MC due to massive evaluation of the \texttt{exp} function whilst computing
the likelihood ratios. Second, the computational burden of NIS increases
sub-linearly for our sample sizes.  
This is due to initial computation for the LBFP, which is
roughly independent of $N$. Remarkably, NIS is computationally cheaper than CDIS 
for {$N=$ 10,000}.

\noindent
\textbf{Example 3.} The last example is a two-dimensional benchmark integration
problem discussed in Givens and Raftery (1996). The density of interest $p(x_1,
x_2)$ is given by $X_1 \sim \mathcal{U}[-1,4]$ and $X_2|X_1 \sim \mathcal{N}(|X_1|, .09a^2)$. 
We investigate the cases $a=.75$ and $a=3.5$. 
This kind of density also occurs in work on whale modeling (Raftery 1995). Small
values for $a$ imply a strong nonlinear dependency between $X_1$ and $X_2$.
As $a$ becomes larger the dependency vanishes in favor of a more diffuse
relationship (see Figure \ref{fig:bayesian}). 
Following Givens and Raftery (1996), we use this scenario for comparing self-normalized IS
algorithms. NSIS is tested against SIS with proposal equal to the
uniform distribution on $[-4,7] \times [-4,8]$. 
The same uniform distribution is used as trial distribution
$q_0$ in the NSIS algorithm.
We compute the expectation of functions $\varphi_1(x_1, x_2) = x_2$ and
$\varphi_2(x_1, x_2) = \mathbf{1}_{\{x_1 < 0 \}}(x_1, x_2)$. 
The parameters of NSIS are set as follows: $\lambda=.2$ and $h=1.54$, $1.224$,
$1.09$ (for {$N=$ 1,250}, 5,000, 10,000). For comparison, we also state the
results of two other nonparametric algorithms, namely GAIS and LAIS (West
1992; Givens and Raftery 1996). GAIS and LAIS are adaptive nonparametric IS
methods, that  estimate distribution $p$ with adaptive envelope refinements based on nonparamtric
kernel estimators.
Density $p$ and the optimal SIS proposals are shown in Figure
\ref{fig:bayesian}. They are rather far away from the initial guess $q_0$. 
Table \ref{table:bayesian} shows the relative efficiency of NSIS, GAIS, and
LAIS with respect to SIS for the two functions and the two different values of
$a$. The figures for GAIS and LAIS were reprocessed from Givens and Raftery
(1996). For {$N=$ 5,000}, NSIS is clearly the method of choice.

\begin{figure}

\centering
\includegraphics[
     height=216pt, width=432pt, keepaspectratio]
     {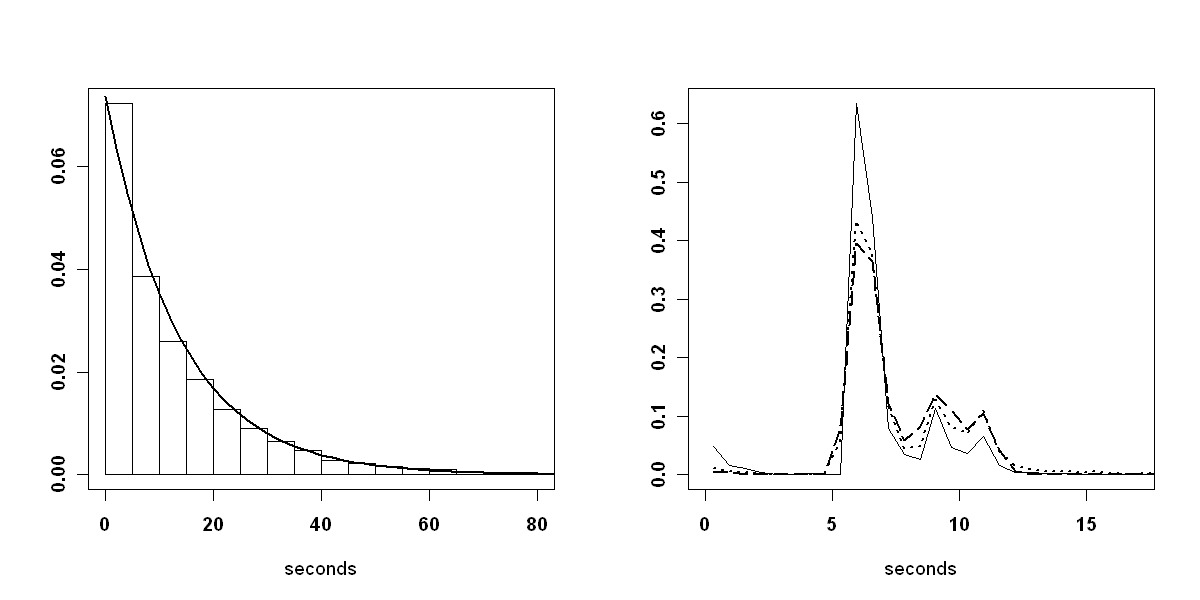}
\caption{\footnotesize Spam filter application: Histogram of the empirical
interarrival times and Exponential distribution with parameter $.074$ (left).
Linear blend frequency polygon estimates of the service time distribution 
(solid line) and of the optimal proposal $q_{\varphi}^{\text{opt}}$ for the
single server (dotted line) and dual server (dashed line) case for $K=10$
(right).}
\label{fig:spamdistributions}

\centering
\includegraphics[
     height=216pt, width=216pt, keepaspectratio]
     {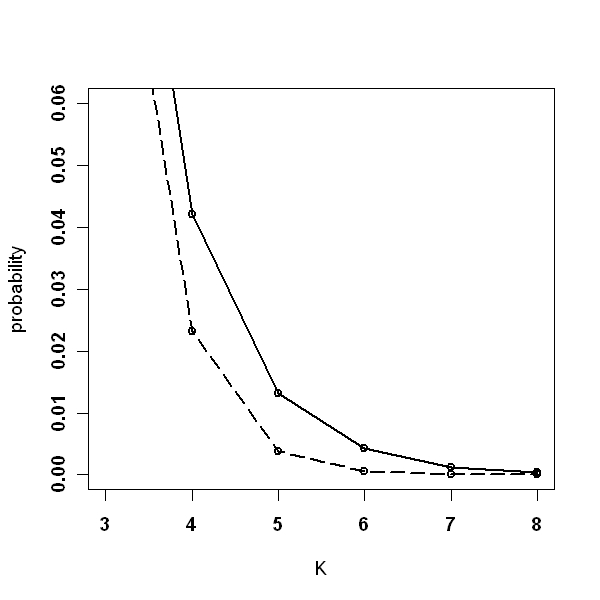}
\caption{\footnotesize Results for spam filter application: Estimated
probabilities of the queue length to reach level $K$ for single server (heavy
line) and dual server (dashed line) case.}
\label{fig:spamprobs}

\end{figure}

\begin{center}
6. APPLICATION
\end{center}
We investigate a spam filter queueing systems with real data.
Queueing system are an active field of research (see for instance Lazowska
1984; Asmussen 2003). Numerous applications are readily available. The most
basic queueing system, denoted briefly by M|M|1, consists of a single server and a single waiting room
(with infinite capacity). The interarrival and service times of the jobs are
exponential distributed with parameter $\mu$ and $\nu$, respectively. This
model is well understood theoretically but usually too restrictive for real
world applications. In our case, e-mails arrive at a spam filter that decides
whether or not a particular e-mail is spam. 
The data consist of interarrival times $t_i$
(in seconds) and service times $s_i$ (in milliseconds) for $n=22,248$ e-mails.
The data were recorded between 8am and 8pm on 8
business days in September 2008 and are available on request.
(We are grateful to J. Kunkel for providing the data.)
The system that produced the data is a single queue, dual server system, i.e. 
the e-mails are processed by two parallel spam filter threads.
In the following we
investigate both the single and the dual server case.
The empirical distributions of the interarrival and service times are displayed
in Figure \ref{fig:spamdistributions}. 
We observe that the former is well approximated by an
exponential distribution with parameter $\hat{\mu} = n/\sum_{i=1}^n t_i = .074$
(which is the maximum likelihood estimate). On the contrary, for the
service time distribution it is hard to find a parametric model. 
Therefore we employ a LBFP estimate.
(Note that a kernel estimator is inappropriate as heavy sampling from the
service time distribution is required.)
The bin width was selected with the Gaussian reference rule for frequency
polygons $\hat{h}=2.15 \hat{\sigma} n^{-1/5}$ (Terrell and Scott 1985) with
$\hat{\sigma}$ being the standard deviation of the service times $s_i$.

We are interested in the probability that the queue length reaches a certain
level $K$. This is a typical problem in queueing systems with rare events being
of particular interest. Importance sampling is a standard variance
reduction technique for this task (see for instance Glynn and Iglehart 1989;
Glasserman and Kou 1995; Kim et al. 2000). 
For estimating the probabilities we simulate $N$ busy periods and count the
number of periods in which level $K$ was reached. 
A busy period begins when an e-mail has arrived in an empty system and ends
when either the system is empty again or the queue length has reached level $K$.
Let $\omega_i$ be the sample path of the queue length in the $i$th busy period
resulting from samples $\mathbf{x}_i^j$ and $\mathbf{y}_i^k$ drawn from the
interarrival distribution $p_t$ and service time distribution $p_s$,
respectively. 
In the dual server case $\mathbf{y}_i^k$ represent the service times of both
servers.
The MC estimate of the probability of interest is
$\hat{I}_K = 1/N \sum_{i=1}^{N} \varphi(\omega_i)$
where $\varphi(\omega_i) = 1$ if $\omega_i$ reaches $K$ and $0$ else.
Assume the number of e-mails that have been served in the $i$th busy
period is $L_i$. Then there must be $K+L_i-1$ arrivals in this period for
the queue to reach level $K$. (Note, a busy period starts with one job in the
queue.) Hence, if importance sampling is used the estimator becomes
\begin{equation*}  
\hat{I}_K^{\text{IS}} = \frac1N \sum_{i=1}^{N} \varphi(\omega_i) l(\omega_i)
\end{equation*}
with likelihood ratio
\begin{equation*}  
l(\omega_i) = \prod_{j=1}^{K+L_i-1}
\frac{p_t(\mathbf{x}_i^j)}{q_t(\mathbf{x}_i^j)}
\prod_{k=1}^{L_i} \frac{p_s(\mathbf{y}_i^k)}{q_s(\mathbf{y}_i^k)}
\end{equation*}
and proposals $q_t$, $q_s$.
Here NIS works as follows: 
We simulate $M$ busy periods by sampling interarrival times
$\tilde{\mathbf{x}}_i^j$ and service times $\tilde{\mathbf{y}}_i^k$ from trial
distributions $q_{0,t}$ and $q_{0,s}$, respectively, and obtain sample paths
$\tilde{\omega}_i$, $i=1,\ldots,M$.  Let $\mathcal{I} = \{i \in \{1,\ldots,M\},
\varphi(\tilde{\omega}_i)=1 \}$.
For estimation of the optimal proposals we use those
times $\tilde{\mathbf{x}}_i^j$, $\tilde{\mathbf{y}}_i^k$ with $i \in
\mathcal{I}$.
The interarrival time proposal $\hat{q}_t$ is estimated parametrically by using
an exponential distribution with parameter
\begin{equation}\label{spamParametricProposal}
\hat{\mu} = \sum_{i \in \mathcal{I}} \sum_{j=1}^{K+\tilde{L}_i-1}
w_i^j
/ \sum_{i \in \mathcal{I}} \sum_{j=1}^{K+\tilde{L}_i-1}
w_i^j \tilde{\mathbf{x}}_i^j
\end{equation}
where $w_i^j = p_t(\tilde{\mathbf{x}}_i^j) /
q_{0,t}(\tilde{\mathbf{x}}_i^j)$.
The service time proposal $\hat{q}_s$ is estimated nonparametrically (as in
Algorithm 1) based on samples $\tilde{\mathbf{y}}_i^k$ and weights $w_i^k
= p_s(\tilde{\mathbf{y}}_i^k) / q_{0,s}(\tilde{\mathbf{y}}_i^k)$, $i \in
\mathcal{I}$.

For our simulation we set $N=$ 1 Mio.\ , $\lambda = .15$, and the
trial distribution $q_{0,s}$ equal to the LBFP estimate of the service
distribution. For M|M|1 systems it is well known that (asymptotically) optimal
proposals are achieved by swapping the parameters $\mu$ and $\nu$. For this reason
$q_{0,t}$ is set to the Exponential distribution with parameter
$\hat{\nu} = n/\sum_{i=1}^n s_i = 0.147$.
As parametric IS benchmark we consider the IS scheme that carries out IS for
the interarrival times only. It uses the Exponential distribution with
parameter $\hat{\mu}$ defined in (\ref{spamParametricProposal}) as proposal.

We compare MC, IS, and NIS in terms of the coefficient of variation (CV) and
RE. The former is defined as the ratio of the standard deviation to the mean of
the probability estimate. Note that for CV smaller values are favourable.
The results are summarized in
Tables \ref{table:spamfiltersingleserver} and \ref{table:spamfilterdualserver}.
Where no figure is given, the MC estimator was zero.
We find that as the event of interest becomes rarer NIS becomes
more favourable. This holds for both the single and dual server case.
The NIS probability estimates for different queue
levels $K$ are shown in Figure \ref{fig:spamprobs}. No error bounds are given
as they are very small for the large number of busy periods used.

Real-world queueing systems typically involve complicated
distributions such as the service time distribution in our case. Therefore, it
is often impossible to set up parametric IS schemes for
simulation. Here, NIS has a distinct advantage.
The extension of NIS to the recently proposed state-dependent IS schemes for
queueing systems is part of our current research.

\begin{center}
7. CONCLUDING REMARKS
\end{center}
Contrary to other articles on nonparametric importance sampling, we favored
the LBFP instead of kernel estimators. 
As shown in Section 4, draws from a LBFP can be generated using
the inversion method. 
As the inversion method is a monotone transformation, it preserves the structure
of the pre-sampled uniformly distributed variates.
This offers the opportunity to combine NIS/NSIS with other variance
reduction techniques such as stratified sampling, moment matching, and quasi
MC techniques (Robert and Casella 2004; Glasserman 2004).

In financial engineering and many other fields, integration problems are
often high-dimensional.
Due to the curse of dimensionality and increasing computational complexity,
the direct application of NIS is intractable for large dimensions. However,
dimension reduction techniques such as principal component analysis, the
Brownian Bridge, or the screening method can
be applied to break down the required integration task to moderate dimensions
(Glasserman 2004; Rubinstein 2007).

Furthermore, we emphasize that the LBFP estimator is not restricted
to the usage within nonparametric importance sampling. It is a reasonable 
alternative to other nonparametric estimators whenever sampling and evaluation
is required.

\begin{center}
APPENDIX A
\end{center}

\noindent
{\bf Proof of Theorem \ref{thm:nis}.}
We denote $q_M^{\text{IS}}$ and
$\hat{q}_M^{\text{IS}}$ briefly by $q_M$ and $\hat{q}_M$. 
Since for $\varphi \geq
0$ we have $q_M = \varphi_M p I_{\varphi_M}^{-1}$, the variance $\sigma_M^2$
of $\hat{I}_{\varphi_M}^{\text{NIS}}$ (conditional on $\{\tilde{\mathbf{x}}^1,
\ldots, \tilde{\mathbf{x}}^M \}$) is given by
\begin{equation}\label{proof:nis:initvar}
(N-M) \sigma^2_M = I_{\varphi_M}^2 \int \frac{(\hat{q}_M(\mathbf{x}) -
q_M(\mathbf{x}))^2}  {\hat{q}_M(\mathbf{x})} d\mathbf{x}.
\end{equation}
In order to get rid of $\hat{q}_M(\mathbf{x})$ in the denominator we write
\begin{eqnarray*}
\frac{N-M}{I_{\varphi_M}^2} \ \mathbf{E} [\sigma^2_M] &=&  \mathbf{E} \left[\int
\frac{(\hat{q}_M(\mathbf{x}) - q_M(\mathbf{x}))^2} {q_M(\mathbf{x})}
d\mathbf{x} \right] - \mathbf{E} \left[\int
\frac{(\hat{q}_M(\mathbf{x}) - q_M(\mathbf{x}))^3} {\hat{q}_M(\mathbf{x})
q_M(\mathbf{x})} d\mathbf{x} \right]\\
&=& K_M + R_M.
\end{eqnarray*}
The discrepancy between $\hat{q}_M$ and $q_M$ can be investigated by
\begin{eqnarray}
\nonumber \hat{q}_M(\mathbf{x}) - q_M(\mathbf{x}) &=&
\frac{\hat{f}_M(\mathbf{x}) - \overline{\omega}_M q_M(\mathbf{x})}
{I_{\varphi_M}} + \ \frac{\delta_M (1- V_M q_M(\mathbf{x}))}
{\overline{\omega}_M + V_M \delta_M}\\ \nonumber && \; + \left[
\frac{\hat{f}_M(\mathbf{x}) - \overline{\omega}_M q_M(\mathbf{x})} {I_{\varphi_M}} \right] \left(
\frac{I_{\varphi_M}} {\overline{\omega}_M + V_M \delta_M} -1 \right)\\
\label{thm:nis:qqhat} &=& W_M(\mathbf{x}) + U^1_M(\mathbf{x}) +
U^2_M(\mathbf{x}).
\end{eqnarray}
It will be established below that $\mathbf{E}[W_M(\mathbf{x})]^2 = O(h^4 +
(Mh^d)^{-1})$.  Now we show that $\mathbf{E}[U^1_M(\mathbf{x}) +
U^2_M(\mathbf{x})]^2$ is of lower order.
Under {Assumptions 1, 2} we yield
\begin{eqnarray*}
\lefteqn{\mathbf{E}[U^1_M(\mathbf{x}) + U^2_M(\mathbf{x})]^2} \\ 
&\leq& C (V_M\delta_M)^2 + C \left( \mathbf{E}\left[ 
\frac{\hat{f}_M(\mathbf{x}) - \overline{\omega}_M q_M(\mathbf{x})}
{I_{\varphi_M}} \right]^4 \right)^{1/2} \left( \mathbf{E}\left[
\frac{I_{\varphi_M}} {\overline{\omega}_M + V_M \delta_M} -1 \right]^4
\right)^{1/2}\\ &\leq& C (V_M\delta_M)^2 + C \left( \frac{1}{Mh^d} + h^4\right)
\left(
\frac{1}{M^3  (V_M \delta_M)^4} + (V_M \delta_M)^2 + \frac{1} {M^2}
\right)^{1/2}.
\end{eqnarray*}
The last inequality follows analogously to {Lemma 1, 2} in Zhang (1996).
Since by Assumption 3 $V_M\delta_M = o(h^2)$ and $M^3 (V_M \delta_M)^4
\rightarrow \infty$, we obtain
$
\mathbf{E}[U^1_M(\mathbf{x}) + U^2_M(\mathbf{x})]^2 =
o(\mathbf{E}[W_M(\mathbf{x})]^2).
$
We conclude $K_M \approx \int \mathbf{E}[W_M(\mathbf{x})^2] q_M^{-1}(\mathbf{x})
d\mathbf{x}$.

It is not hard to work out that $\int \mathbf{E}[W_M(\mathbf{x})^2] q_M^{-1}(\mathbf{x})
d\mathbf{x}$ decomposes into an integrated
squared bias term $L_1$ and an integrated variance term $L_2$
\begin{equation*}
\int \frac{(\mathbf{E}[\hat{f}_M(\mathbf{x}) I_{\varphi_M}^{-1}] -
q_M(\mathbf{x}))^2} {q_M(\mathbf{x})} d\mathbf{x} + \int
\frac{\text{Var}[\hat{f}_M(\mathbf{x}) I_{\varphi_M}^{-1}]} {q_M(\mathbf{x})}
d\mathbf{x} + O(M^{-1})
= L_1 + L_2 + O(M^{-1}).
\end{equation*}
For notational convenience the following is only shown for $d=1$.
Without loss of generality we assume $\mathbf{x} \in [-h/2, h/2)$. Then
$\hat{f}_M I_{\varphi_M}^{-1}$ simplifies to
\begin{equation}\label{proof:nis:lbfpsimple}
\frac{\hat{f}_M(\mathbf{x})} {I_{\varphi_M}} = \left(\frac{h/2 -
\mathbf{x}}{h}\right)
\frac{\hat{f}^{\text{UH}}_0}{I_{\varphi_M}} + \left(\frac{h/2 +
\mathbf{x}}{h}\right)
\frac{\hat{f}^{\text{UH}}_1}{I_{\varphi_M}}
\end{equation}
where $\hat{f}^{\text{UH}}_0 = 1/M \sum_{j=1}^M \omega^j_M \mathbf{1}_{
[-h, 0)}(\tilde{\mathbf{x}}^j)$ and
$\hat{f}^{\text{UH}}_1 = 1/M \sum_{j=1}^M \omega^j_M \mathbf{1}_{
[0, h)}(\tilde{\mathbf{x}}^j)$ are the heights of bins $[-h, 0)$, respectively,
$[0, h)$ of an unnormalized histogram.
For the computation of $L_1$ we need to compare the Taylor expansions of
$\mathbf{E}[\hat{f}_M(\mathbf{x}) I_{\varphi_M}^{-1}]$ and $q_M$ which are given
by
\begin{eqnarray*}
\mathbf{E}[\hat{f}_M(\mathbf{x}) I_{\varphi_M}^{-1}] 
&=& q_M(0) + \mathbf{x} q_M'(0) + h^2
q_M''(0)/6 + O(h^3),\\
q_M(\mathbf{x}) &=& q_M(0) + \mathbf{x} q_M'(0) + \mathbf{x}^2
q_M''(0)/2 + O(h^3).
\end{eqnarray*}
The former follows from (\ref{proof:nis:lbfpsimple}) and from the
expansion of the histogram
\begin{equation*}
\mathbf{E}[\hat{f}^{\text{UH}}_{0/1} I_{\varphi_M}^{-1}] = q_M(0) \ {-/+} \ h
q_M'(0)/2 + h^2 q_M''(0)/6 + O(h^3).
\end{equation*}
Thus we obtain $(\mathbf{E}[\hat{f}_M(\mathbf{x}) I_{\varphi_M}^{-1}]
 - q_M(\mathbf{x}))^2 \approx (h^2 - 3 \mathbf{x}^2)^2 q_M''(0)^2/36 $.
Integration over $[-h/2, h/2)$ and using Taylor expansion of $1/q_M(\mathbf{x})$
about $0$ leads to
\begin{equation}\label{thm:nis:prebias2}
\frac{q_M''(0)^2}{36} \int_{-h/2}^{h/2} \frac{(h^2 -3
\mathbf{x}^2)^2}{q_M(\mathbf{x})} d\mathbf{x} = \frac{49 \ q_M''(0)^2}{2880 \
q_M(0)} h^5 + O(h^6).
\end{equation}
By summing over all bins and applying standard Riemann approximation we yield
\begin{equation}\label{thm:nis:bias2}
L_1 = \frac{49}{2880} h^4 \int \frac{q_M''(\mathbf{x})^2}{q_M(\mathbf{x})}
d\mathbf{x} + O(h^5).
\end{equation}
Next we derive an approximation to $L_2$. From (\ref{proof:nis:lbfpsimple}) we
have
\begin{eqnarray*}
\text{Var}[\hat{f}_M(\mathbf{x}) I_{\varphi_M}^{-1}] &=&
\left(\frac{h/2-\mathbf{x}} {h}\right)^2
\text{Var}[\hat{f}^{\text{UH}}_0 I_{\varphi_M}^{-1} ] +
\left(\frac{h/2+\mathbf{x}} {h}\right)^2
\text{Var}[\hat{f}^{\text{UH}}_1 I_{\varphi_M}^{-1}] \\
&& \; + \; \frac{h^2/2- 2
\mathbf{x}^2} {h^2} \text{Cov}[\hat{f}^{\text{UH}}_0 I_{\varphi_M}^{-1}, \hat{f}^{\text{UH}}_1 I_{\varphi_M}^{-1}].
\end{eqnarray*}
In addition, it can be shown that $\text{Var}[\hat{f}^{\text{UH}}_i
I_{\varphi_M}^{-1}] \approx \frac{q_M(0)^2} {M h q_0(0)} - \frac{q_M(0)^2}{M}$ for $i=0,1$ and
$\text{Cov}[\hat{f}^{\text{UH}}_0 I_{\varphi_M}^{-1}, \hat{f}^{\text{UH}}_1
I_{\varphi_M}^{-1}] \approx -\frac{q_M(0)^2} {M}$ analogously to Scott (1992, chap. 4).
That is we yield
\[\text{Var}[\hat{f}_M(\mathbf{x}) I_{\varphi_M}^{-1}] = \left(
\frac{1}{2Mh} + \frac{2\mathbf{x}^2}{Mh^3} \right)\frac{q_M(0)^2}
{q_0(0)} + O(M^{-1}).\]
Analogously to (\ref{thm:nis:prebias2}) and
(\ref{thm:nis:bias2}) we then obtain 
\begin{equation*}
\int_{-h/2}^{h/2} \frac{\text{Var}[\hat{f}_M(\mathbf{x}) I_{\varphi_M}^{-1}]}
{q_M(\mathbf{x})} d\mathbf{x} = \frac{2 q_M(0)} {3M q_0(0)} + O(h/M)
\end{equation*}
and
\begin{equation*}
L_2 = \frac{2}{3Mh} \int \frac{q_M(\mathbf{x})} {q_0(\mathbf{x})} d\mathbf{x}
+ O(M^{-1}),
\end{equation*}
respectively.
Very similar computations in the multivariate case yield
\[K_M \approx h^4 H_{M,1} + \frac{2^d}{3^d M h^d} H_{M,2}\] where
\begin{equation*}
H_{M,1} = \frac{49}{2880} \sum_{i=1}^{d} \int
\frac{(\partial_i^2 q_M)^2}{q_M}+ \frac{1}{64} \sum_{i \not= j} \int \frac{\partial_i^2
q_M \partial_j^2 q_M}{q_M} \quad \text{and} \quad H_{M,2}= \int
\frac{q_M}{q_0}.
\end{equation*}
It remains to show that $R_M$ is negligible compared to
$K_M$. The construction of $\hat{g}_M$ implies $\hat{g}_M \geq
\delta_M (\overline{\omega}_M + V_M \delta_M)^{-1} > 0$. Under {Assumption 1}
and 4a it can be shown that $R_M$ is negligible following
the same lines as in Zhang (1996).

The proof is finished by noting that the squared bias term in
$\mathbf{E}[\hat{I}_{\varphi_M}^{\text{NIS}} - I_{\varphi}]^2$
is negligible due to {Assumption 5a} and that the expressions
$I_{\varphi_M}^2$ (in (\ref{proof:nis:initvar})), $H_{M,1}$, and $H_{M,2}$ can
be substituted by their unrestricted counterparts as their differences are of lower order.

\medskip
\noindent
{\bf Proof of Theorem \ref{thm:nis2}.}
Again $q_M$ is shorthand for $q_M^{\text{IS}}$. Let  $f_{\varphi_M} =
\left(\frac{\varphi_M p}{I_{\varphi_M}} - \frac{|\varphi_M|
p}{\overline{I}_{\varphi_M}}\right)$. 
Straight forward calculations yield
\begin{eqnarray*}
(N-M) \sigma^2_M &=& I_{\varphi_M}^2 \int \left( \frac{\varphi_M
p}{I_{\varphi_M}} - \frac{ |\varphi_M| p}{\overline{I}_{\varphi_M}} + q_M -\hat{q}_M \right)^2
\hat{q}_M^{-1}\\ 
&=& I_{\varphi_M}^2 \left[ \int f_{\varphi_M}^2
\frac{(q_M - \hat{q}_M)}{q_M \hat{q}_M} + 2\int f_{\varphi_M} 
\frac{(q_M - \hat{q}_M)}{\hat{q}_M} + \int \frac{(q_M - \hat{q}_M)^2}{\hat{q}_M}
+ \int \frac{f_{\varphi_M}^2}{q_M}\right]\\ 
&=& I_{\varphi_M}^2
\left[ T_1 + T_2 + T_3 + T_4 \right].
\end{eqnarray*}
Term $T_4$ is independent of the nonparametric
estimation and we have $I_{\varphi_M}^2 T_4 = \overline{I}_{\varphi_M}^2 -
I_{\varphi_M}^2$. The expectation of term $T_1$ can be written as
\begin{equation*}
\int f_{\varphi_M}^2 \frac{\mathbf{E}[q_M - \hat{q}_M]}{q_M^2} - 
\int f_{\varphi_M}^2 \frac{\mathbf{E}[q_M - \hat{q}_M]^2}{q_M^3} + 
\int f_{\varphi_M}^2 \frac{\mathbf{E}[q_M - \hat{q}_M]^3}{q_M^3 \hat{q}_M} =
T_{1,1} + T_{1,2} + T_{1,3}.
\end{equation*}
Similar expressions are obtained for quantities $T_2$ and $T_3$. 
We begin with $T_{1,1}$.
Analogously to (\ref{thm:nis:qqhat}) we conclude $
q_M(\mathbf{x})-\hat{q}_M(\mathbf{x}) \approx - [\hat{f}_M(\mathbf{x}) -
\overline{\omega}_M q_M(\mathbf{x})] / \overline{I}_{\varphi_M}
$.
From the proof of {Theorem \ref{thm:nis}} we also know that
\begin{equation*}
\mathbf{E}[\hat{f}_M(\mathbf{x}) -
\overline{\omega}_M q_M(\mathbf{x}) / \overline{I}_{\varphi_M}]
= \mathbf{E}[\hat{f}_M(\mathbf{x}) \overline{I}_{\varphi_M}^{-1}] -
q_M(\mathbf{x}) = (h^2 - 3\mathbf{x}^2) q_M''(0)/6 + O(h^3)
\end{equation*}
for $d=1$ and $\mathbf{x} \in [-h/2, h/2)$. Then we obtain
\begin{equation*}
\frac{q_M''(0)}{6} \int_{-h/2}^{h/2} f_{\varphi_M}(\mathbf{x})^2 \frac{h^2 -
3\mathbf{x}^2} {q_M(\mathbf{x})^2} d\mathbf{x} = \frac{h^3}{8}
f_{\varphi_M}(0)^2 \frac{q_M''(0)} {q_M(0)^2} + O(h^4)
\end{equation*}
using a Taylor expansion of $f_{\varphi_M}(\mathbf{x})^2/q_M(\mathbf{x})^2$
about $0$. Finally, summing over all bins and using Riemann approximation gives
$T_{1,1}$ in the one-dimensional case
\begin{equation*}
-\frac{h^2}{8} \int f_{\varphi_M}(\mathbf{x})^2
\frac{q_M''(\mathbf{x})} {q_M(\mathbf{x})^2}  d \mathbf{x} + O(h^3).
\end{equation*}
In the multivariate case we yield
\begin{equation*}
T_{1,1} = -\frac{h^2}{8}
\int f_{\varphi_M}^2(\mathbf{x}) \frac{\nabla^2 q_M(\mathbf{x})}
{q_M(\mathbf{x})^2} d\mathbf{x} + O(h^3).
\end{equation*}
Term $T_{1,2}$ can be treated
analogously to $\int \mathbf{E}[W_M(\mathbf{x})^2] q_M^{-1}(\mathbf{x})
d\mathbf{x}$ in the proof of {Theorem \ref{thm:nis}}.
We end up with
\begin{equation*}
T_{1,2} = - \frac{2^d}{3^d M h^d} \int \frac{f^2_{\varphi_M}}{q_0 q_M} - 
 \left[ \frac{49h^4}{2880} \sum_{i=1}^{d} \int f^2_{\varphi_M}
\frac{(\partial_i^2 q_M)^2}{q_M^3}+ \frac{h^4}{64} \sum_{i \not= j} \int
f^2_{\varphi_M} \frac{\partial_i^2 q_M \partial_j^2 q_M}{q_M^3}\right].
\end{equation*}
Comparing the term in brackets to $T_{1,1}$ we observe that the former is
negligible.
Furthermore, similarly to $R_M$ in the proof of {Theorem \ref{thm:nis}}, it
follows that $T_{1,3}$ is negligible compared to $T_{1,2}$ provided that
{Assumption 4} holds.

The calculations for $T_2$ and $T_3$ are very similar to those of $T_1$ and
therefore omitted. Putting all terms together we obtain
\begin{eqnarray*}
(N-M) \mathbf{E}[\sigma^2_M] &=& I_{\varphi_M}^2 \Bigl\lbrace \frac{2^d}{3^d M
h^d} \left( \int \frac{q_M}{q_0} - 2 \int \frac{f_{\varphi_M}}{q_0} - \int
\frac{f^2_{\varphi_M}}{q_0 q_M} \right)\\
&& \; - h^2\left( \int
f_{\varphi_M}^2 \frac{\nabla^2 q_M}{8 q_M^2} + \int
f_{\varphi} \frac{\nabla^2 q_M}{4 q_M} \right) \Bigr\rbrace
 \times \ (1 + o(1)) 
+ (\overline{I}_{\varphi_M}^2 - I_{\varphi_M}^2).
\end{eqnarray*}
We observe that the terms restricted on $M$ can be substituted by their
asymptotic limits, which completes the proof.

\medskip
\noindent
{\bf Proof of Theorem \ref{thm:nsis}.} 
We denote $q_M^{\text{SIS}}$ and
$\hat{q}_M^{\text{SIS}}$ briefly by $q_M$ and $\hat{q}_M$. 
As the bias of
$\hat{I}_{\varphi_M}^{\text{NSIS}}$ is asymptotically negligible we have
$\mathbf{E}[\hat{I}_{\varphi_M}^{\text{NSIS}} - I_{\varphi}]^2 = (N-M)^{-1} 
\mathbf{E}[\sigma_{\text{SIS}}^2] \times \{1+o(1)\}$. 
Thus, it suffices to examine $\mathbf{E}[\sigma_{\text{SIS}}^2]$ with
$\sigma_{\text{SIS}}^2$ as in (\ref{nnis:variance}). We obtain analogously to
(\ref{thm:nis:qqhat})
\begin{equation*}
\nonumber  \hat{q}_M(\mathbf{x}) - q_M(\mathbf{x}) =
\frac{\hat{f}_M(\mathbf{x}) - \overline{\omega}_M q_M(\mathbf{x})}
{\alpha \tilde{I}_{\varphi_M}} + \widetilde{U}^1_M(\mathbf{x}) +
\widetilde{U}^2_M(\mathbf{x}).
\end{equation*}
The crucial step for proving that the remainder
term $\widetilde{U}^1_M(\mathbf{x}) + \widetilde{U}^2_M(\mathbf{x})$ is of
lower order is to show that under {Assumptions 1, 2}
\[
\mathbf{E}[\alpha \tilde{I}_{\varphi_M} - \overline{\omega}_M]^{2l} \leq C M^{-l}
\]
for $l=1,2$ (compare {Lemma 1, 2} in Zhang (1996)). 
We have
\begin{eqnarray*}
|\alpha \tilde{I}_{\varphi_M} - \overline{\omega}_M | 
&\leq& \left| \alpha \tilde{I}_{\varphi_M} - \frac{1}{M} \sum_{j=1}^{M} |\varphi_M(\tilde{\mathbf{x}}^j) - I_{\varphi_M}|
\tilde{p}(\tilde{\mathbf{x}}^j) q_0(\tilde{\mathbf{x}}^j)^{-1} \right|\\
&& \quad + \frac{1}{M} \sum_{j=1}^{M} \tilde{p}(\tilde{\mathbf{x}}^j)
q_0(\tilde{\mathbf{x}}^j)^{-1} | I_{\varphi_M} - \breve{I}_{\varphi_M} |
\end{eqnarray*}
and by applying the Minkowski inequality we obtain
\begin{eqnarray*}
\left( \mathbf{E}[\alpha \tilde{I}_{\varphi_M} - \overline{\omega}_M]^{2l}
\right)^{\frac{1}{2l}} 
&\leq& \left( \mathbf{E} \left[ \alpha \tilde{I}_{\varphi_M} - \frac{1}{M}
\sum_{j=1}^{M} |\varphi_M(\tilde{\mathbf{x}}^j) - I_{\varphi_M}|
\tilde{p}(\tilde{\mathbf{x}}^j) q_0(\tilde{\mathbf{x}}^j)^{-1} \right]^{2l}
\right)^{\frac{1}{2l}} \\
&& \quad + \ C \left( \mathbf{E}[ I_{\varphi_M} - \breve{I}_{\varphi_M}]^{2l}
\right)^{\frac{1}{2l}}\\ 
&=& C \left( M^{-1/2} + M^{-1/2}\right).
\end{eqnarray*}
Hence, we conclude that the remainder term is of lower order.
Finally, we need to show that 
\begin{equation*}
\int \mathbf{E}\left[
\left( \frac{\hat{f}_M(\mathbf{x}) - \overline{\omega}_M q_M(\mathbf{x})}
{\alpha \tilde{I}_{\varphi_M}} \right)^2 \hat{q}_M(\mathbf{x})^{-1}
 \right] d\mathbf{x} \approx h^4 H_1 + \frac{2^d}{3^d M h^d} H_2.
\end{equation*}
The main difference to {Theorem \ref{thm:nis}} is the 
dependency of the weights $\widetilde{\omega}_M^j$.
Define $\check{\omega}_M^j = |\varphi_M(\tilde{\mathbf{x}}^j) - I_{\varphi_M}|
\tilde{p}(\tilde{\mathbf{x}}^j) q_0(\tilde{\mathbf{x}}^j)^{-1}$, $j=1,\ldots,M$.
As in the proof of {Theorem \ref{thm:nis}} let $\check{f}^{\text{UH}}_{0/1}$
and $\hat{f}^{\text{UH}}_{0/1}$ be unnormalized histogram bins
based on weights $\check{\omega}_M^j$ and $\widetilde{\omega}_M^j$, respectively. 
It is not hard to show that 
$
\mathbf{E} [\hat{f}^{\text{UH}}_{0/1} (\alpha \tilde{I}_{\varphi_M})^{-1}] = 
\mathbf{E} [\check{f}^{\text{UH}}_{0/1} (\alpha \tilde{I}_{\varphi_M})^{-1}] +
\mathcal{O}(M^{-1/2})
$
and
$
\text{Var} [\hat{f}^{\text{UH}}_{0/1} (\alpha \tilde{I}_{\varphi_M})^{-1}] = 
\text{Var} [\check{f}^{\text{UH}}_{0/1} (\alpha \tilde{I}_{\varphi_M})^{-1}] +
\mathcal{O}(M^{-1}).
$
The rest of the proof follows analogously to {Theorem \ref{thm:nis}}
since weights $\check{\omega}_M^j$ are independent 
and the additional $\mathcal{O}(M^{-1/2})$, $\mathcal{O}(M^{-1})$ terms are
negligible.

\begin{center}
APPENDIX B
\end{center}
Let $B_M$ be the number of bins. It follows that the number of bins in each
marginal space is $\mathcal{O}(B_M^{1/d})$. We begin with the analysis of the
evaluation of a LBFP. Given location $\mathbf{x}$ we need to find the
associated bin mid-points $(t_{k_1}, \ldots, t_{k_d})$ which is of order
$\mathcal{O}(dB_M^{1/d})$. Then formula (\ref{def:LBFP}) can be evaluated which
is $\mathcal{O}(2^d d)$. Now observe $B_M \approx V_M/h^d$ and
$h^{*}=\mathcal{O}(\rho(d)^{1/(d+4)} N^{-1/(d+4)})$ with $\rho(d) = d (2/3)^d$.
By assuming that $h=h^{*}$ we obtain $\mathcal{O}(d B_M^{1/d} + 2^d d)
\approx \mathcal{O}(\rho(d)^{-1/(d+4)} d N^{1/(d+4)} + 2^d d)$ neglecting the
slowly increasing sequence $V_M$.

Sampling from a LBFP consists of the three steps described in Section 4.2.
In Step 1 the marginalized histograms corresponding to the LBFP
$\hat{f}(x_{1:i})$, $i=1,\ldots,d-1$, need to be calculated. This can be done
recursively in $\mathcal{O}(B_M)$. In the second step $\hat{F}$ are to be
computed at all bin mid-points $t_{k_i}$ using
relation (\ref{relation:uniFP-LBFP}). Thus it is required to evaluate
$\hat{f}(x_{1:i-1}, t_{k_i})$, $i=1,\ldots,d$. This consists of searching the
bin mid-points $(t_{k_1}, \ldots, t_{k_{i-1}})$ associated with $x_{1:i-1}$ and
evaluating formula (\ref{def:LBFP}) as we have discussed above. It is
sufficient to do the former once. Thus we end up with $\mathcal{O}(d B_M^{1/d} + 2^d d \times d B_M^{1/d})$ where the latter $d B_M^{1/d}$ is due to the
evaluation of $\hat{F}$ at all $t_{k_i}$ in each marginal dimension. Step 3 has
complexity $\mathcal{O}(d B_M^{1/d})$ as in each marginal dimension the bin
mid-point $t_{k_{i}}$ satisfying $y_i \in [\hat{F}(t_{k_i}|x_{1:i-1}),
\hat{F}(t_{k_i+1}|x_{1:i-1}))$ must be found. Putting all together we yield
$\mathcal{O}( B_M + 2^d d^2 B_M^{1/d})$ the generating one sample. As above we
assume $h=h^{*}$, substitute $B_M \approx V_M/h^{d}$, and omit $V_M$ in order
to derive
$\mathcal{O}( \rho(d)^{-d/(d+4)} N^{d/(d+4)} + 2^d d^2
\rho(d)^{-1/(d+4)} N^{1/(d+4)})$. As Step 1 needs to be carried out only once
and as $\rho(d)^{-1/(d+4)}$ is small compared to $2^d d^2$ we obtain
approximately $\mathcal{O}(2^d d^2 N^{(d+5)/(d+4)})$ for generating $N$
samples. Finally, we remark that $N$ evaluations are negligible compared to
generating $N$ samples.

\singlespace

\vspace{1cm}

 \begin{center}
 REFERENCES
 \end{center}

\begin{description}

\item Asmussen, S. (2003), {\it
Applied Probability and Queues}, New York: Springer.

\item Geweke, J. (1989), ``Bayesian Inference in Econometric Models using Monte
Carlo Integration,'' in {\it Econometrica}, 57, 1317-1339.

\item Givens, G. H., and Raftery, A. E. (1996),  ``Local Adaptive Importance
Sampling for Multivariate Densities With Strong Nonlinear Relationships,''  in
{\it Journal of American Statistical Association}, 91, 132-141.

\item Glasserman, P., and Kou, S.-G. (1995),
``Analysis of an Importance Sampling Estimator for Tandem Queues,'' in {\it
ACM Transactions on Modeling and Computer Simulation}, 5, 22-42.

\item Glasserman, P., Heidelberger, P., and Shahabuddin, P. (1999),
``Asymptotically optimal importance sampling and stratification for pricing
path-dependent options,'' in {\it Mathematical Finance}, 9, 117-152.

\item Glasserman, P. (2004), {\it
Monte Carlo Methods in Financial Engineering}, New York: Springer.

\item Glynn, P. W., and Iglehart, D. L. (1989),  ``Importance Sampling for
Stochastic Simulations,'' in {\it Management Science}, 35, 1367-1392.

\item Kim, Y. B., Roh, D. S., and Lee, M. Y. (2000), ``Nonparametric
Adaptive Importance Sampling For Rare Event Simulation,'' in {\it
Winter Simulation Conference Proceedings}, Vol. 1, 767-772.

\item Kollman, C., Baggerly, K., Cox, D., and Picard, R. (1999),
``Bayesian Inference in Econometric Models using Monte Carlo Integration,'' in
{\it Annals of Applied Probability}, 9, 391-412.

\item Lazowska, E. D. (1984), {\it Quantitative System Performance, Computer
System Analysis Using Queuing Network Models}, Prentice Hall.

\item Matsumoto, M., and Nishimura, T. (1998), ``Mersenne Twister: A
623-Dimensionally Equidistributed Uniform Pseudo-Random Number Generator,'' in
{\it ACM Transactions on Modeling and Computer Simulations}, 8, 3-30.

\item Musso, M., Oudjane, N., and Le Gland, F. (2001), ``Improving Regularised
Particle Filters,'' in {\it Sequential Monte Carlo Methods in Practice}, eds.
A. Doucet, N. de Freitas and N. Gordon, New York: Springer.

\item Oh, M. S., and Berger, J. (1992), ``Adaptive Importance Sampling in Monte
Carlo Integration,'' in {\it Journal of Statistical Computation and Simulation},
41, 143-168.

\item --- (1993), ``Integration of Multimodal Functions
by {Monte} {Carlo} Importance Sampling,'' in {\it Journal of American
Statistical Association}, 88, 450-456.

\item Raftery, A. E., Givens, G. H., and Zeh, J. E. (1995), ``Inference
from a Deterministic Population Dynamics Model for Bowhead Whales,'' in
{\it Journal of American Statistical Association}, 90, 402-430. 

\item Robert, C. P., and Casella, G. (2004), {\it
Monte Carlo Statistical Methods}, New York: Springer.

\item Rubinstein, R. Y. (1981), {\it
Simulation and the Monte Carlo Method}, New York: Wiley.

\item Rubinstein, R. Y. (2007), ``How to Deal with the Curse of
Dimensionality of Likelihood Ratios in Monte Carlo Simulation,'' unpublished
manuscript.

\item Scott, D. W. (1992), {\it
Multivariate Density Estimation}, New York: Wiley.

\item Stadler, J. S., and Roy, S. (1993), ``Adaptive Importance Sampling,'' in
{\it IEEE journal on selected areas in communications}, 11, 309-316.

\item Terrell, G. R., and Scott, D. W. (1985), ``Oversmoothed Nonparametric
Density Estimates,'' in {\it Journal of American Statistical Association}, 80,
209-214.

\item West, M. (1992), ``Modelling with Mixtures,'' in {\it Bayesian Statistics
4}, eds. J.M. Bernardo et al., Oxford UK: Oxford University Press, 503-524.

\item --- (1993), ``Approximating Posterior Distributions by Mixtures,'' in
{\it Journal of Royal Statistical Society}, 55, 409-422.

\item Zhang, P. (1996), ``Nonparametric Importance Sampling,'' in {\it Journal
of American Statistical Association}, 91, 1245-1253.

\item Zlochin, M., and Baram, Y. (2002), ``Efficient nonparametric importance
sampling for Bayesian learning,'' in {\it Neural Networks} 2002, 2498--2502.

\end{description}

\vspace{1cm}

 \begin{center}
 TABLES AND FIGURES
 \end{center}

\begin{table}[h!]
\centering
\footnotesize
\begin{tabular}{lc|cc|cc|cc}
\hline
& & $N=$ 1,000 & & $N=$ 5,000 & & $N=$ 10,000\\
Method & $d$ & RE & Time &  RE & Time & RE & Time \\
\hline
\hline
MC     & 1 & 1.0  & 1  & 1.0  & 9    & 1.0  & 16   \\
IS     & 1 & 1.5  & 3  & 1.8  & 16   & 1.6  & 31   \\
NIS    & 1 & 1.6  & 13 & 1.8  & 28   & 1.7  & 50   \\
NIS+/- & 1 & 25.0 & 13 & 57.3 & 24   & 51.3 & 40   \\
\hline
MC     & 4 & 1.0 & 4   & 1.0  & 20   & 1.0  & 45   \\
IS     & 4 & 5.0 & 7   & 5.2  & 38   & 4.2  & 80   \\
NIS    & 4 & 3.1 & 112 & 4.0  & 234  & 3.8  & 408  \\
NIS+/- & 4 & 9.1 & 105 & 26.0 & 195  & 22.0 & 326  \\
\hline
MC     & 8 & 1.0  & 13  & 1.0  & 60   & 1.0  & 121  \\
IS     & 8 & 18.6 & 20  & 23.0 & 104  & 26.3 & 209  \\
NIS    & 8 & 7.8  & 600 & 17.4  & 1460 & 5.4  & 4020 \\
NIS+/- & 8 & 7.5  & 572 & 30.2 & 1290 & 37.4 & 2170 \\
\hline
\end{tabular}
\caption{\footnotesize Simulation results for Example 1. 
All figures are computed/averaged over 100 independent runs.}
\label{table:posneg}
\end{table}

\begin{table}
\centering
\footnotesize
\begin{tabular}{l|ccc}
\hline
& $N=$ 1,000 & $N=$ 5,000 & $N=$ 10,000\\
Method & Time (ms) & Time (ms) & Time (ms) \\
\hline
\hline
MC   & 1.8  & 9.0  & 17.8 \\
CDIS & 6.0  & 27.8 & 54.5 \\
NIS  & 13.7 & 29.2 & 48.9 \\
NSIS & 14.1 & 31.1 & 52.1 \\
\hline
\end{tabular}
\caption{\footnotesize CPU times for the option pricing example (Example 2)
averaged over 1,000 independent runs.}
\label{table:optionPricingTimes}
\end{table}

\begin{table}
\centering
\footnotesize
\begin{tabular}{lc|cc|cc}
\hline
& & $\varphi_1$ & & $\varphi_2$ & \\
Method & $N$ & $a=.75$ & $a=3.5$ & $a=.75$ & $a=3.5$\\
\hline
\hline
NSIS & 1,250 & 1.59 & 2.89 & 0.58 & 3.82\\
GAIS & 1,250 & 0.02 & 3.45 & 0.30 & 1.11\\
LAIS & 1,250 & 0.75 & 0.99 & 1.92 & 0.58\\
\hline
NSIS & 5,000 & 8.08 & 4.50 & 9.21 & 5.09\\
GAIS & 5,000 & 5.88 & 0.67 & 0.96 & 0.36\\
LAIS & 5,000 & 3.45 & 1.30 & 2.63 & 0.42\\
\hline
NSIS & 10,000 & 9.38 & 4.75 & 11.06 & 5.77\\
\hline
\end{tabular}
\caption{\footnotesize Relative efficiency of NSIS, GAIS, and LAIS compared to
SIS for Example 3. Figures for NSIS are computed over 1,000 independent runs.
Figures for GAIS and LAIS are reprocessed from Table 2 in Givens and Raftery
(1996).}
\label{table:bayesian}
\end{table}

\begin{table}[h!]
\centering
\footnotesize
\begin{tabular}{l|cc|cc|cc|cc}
\hline
& $K=$ 5 & & $K=$ 10 & & $K=$ 20 & & $K=$ 30 &\\
Method & RE & CV & RE & CV & RE & CV & RE & CV \\
\hline
\hline
MC     & 1.0  & .001  & 1.0  & .14    & -  & -  & -  & -   \\
IS     & .3   & .01 & 19.8  & .03   & -  & .08   & -  & .24   \\
NIS    & .2   & .02 & 58.4  & .02   & -  & .03   & -  & .09   \\
\hline
\end{tabular}
\caption{\footnotesize Results for the spam filter queueing
application (single server case). 
Relative efficiency and coefficient of variation for the estimates of the
probability that the queue length reaches level $K$.
All figures are computed over 100 independent runs with 1 Mio.\ busy periods in
each run.}
\label{table:spamfiltersingleserver}
\end{table}

\begin{table}[h!]
\centering
\footnotesize
\begin{tabular}{l|cc|cc|cc}
\hline
& $K=$ 4 & & $K=$ 6 & & $K=$ 8\\
Method & RE & CV & RE & CV & RE & CV\\
\hline
\hline
MC     & 1.0  & .007  & 1.0  & .044    & 1.0  & .34\\
IS     & 3.7   & .003 & 7.0  & .017   & 53.5  & .046\\
NIS    & 2.6   & .004 & 24.3  & .009   & 184.6  & .025\\
\hline
\end{tabular}
\caption{\footnotesize Results for the spam filter queueing
application (dual server case). 
Relative efficiency and coefficient of variation for the estimates of the
probability that the queue length reaches level $K$.
All figures are computed over 100 independent runs with 1 Mio.\ busy periods in
each run.}
\label{table:spamfilterdualserver}
\end{table}

\end{document}